# Experience with the Hubble Space Telescope: Twenty Years of an Archetype

Matthew D. Lallo

Space Telescope Science Institute

3700 San Martin Dr., Baltimore, MD, 21218

lallo@stsci.edu

**Abstract**:

The Hubble Space Telescope's mission is summarized, with special emphasis placed on the Space Telescope Science Institute's unique experience with Hubble's behavior as an astronomical telescope in the environment of low earth orbit for over two decades. Historical context and background are given, and the project's early scientific expectations are described. A general overview of the spacecraft is followed by a more detailed look at the optical design, both as intended and as built. Basic characteristics of the complete complement of Science Instruments are also summarized. Next our experience with the telescope on-orbit is reviewed, starting with the major initial problems, the solutions, the human servicing missions, and the associated expansion of the observatory's capabilities over this time. Specific attention is then given to our understanding of Hubble's optical quality and pointing/jitter performance, two fundamental characteristics of a telescope. Experience with–and the important mitigation of– radiation damage and contamination is also related. Beyond the telescope itself, we briefly discuss the advances in data reduction, calibration, and observing techniques, as well as the subsequent emergence of highly accessible high-level archival science products. The paper concludes with Hubble's scientific impact.

# 1. INTRODUCTION

The Hubble Space Telescope is now quite possibly the most well known scientific instrument in history, with recognition by the general public throughout the world. Its name alone now conjures images of the spectacular. It has become a symbol for ingenuity in our quest for understanding. It is a source of pride not just among Americans or Europeans, but humans. This was not always the case. Its early gestation was riddled with budget and schedule crises and technical difficulties; when it finally did make it to orbit, years late and roughly double in price, it quickly became notorious as a costly mistake. The disappointment of the early years, however, fades in the glow of Hubble's successes. But beyond the highs and lows of this mission, and the hyperbole often used in both cases, there is simply a very intelligently designed, well-engineered, and well cared for '70s era telescope, protected within an aging yet remarkably stable and robust spacecraft, and sporting state of the art science instruments. The flexibility and extensibility designed into this unique observatory, combined with the dramatic periodic refurbishments in orbit and creative new ways to use it have resulted in a number of increasingly capable incarnations. With Hubble we have learned about the universe, but of particular interest to this audience is also what Hubble's enduring mission teaches us about observatories in space.

## 2. EARLY PROJECT HISTORY

The notion of an astronomical telescope in space had existed since the 1920s, but the heritage of the modern general-purpose "Large Space Telescope" can be traced back to the seminal work and career-long advocacy of Lyman Spitzer [1][2][3]. Throughout much of the 1970s, the case for such a telescope and its concomitant funding was slowly but convincingly built, and by 1977 funding for the design and development of "The Space Telescope" (ST) was appropriated by congress, with 15% of the costs being assumed by the European Space Agency (ESA) in the form of various flight hardware and staffing contributions.

By 1979 construction of a large (2-meter class) human-serviceable space observatory was well underway, with Lockheed Missiles and Space Company acting as the Prime Contractor responsible for integration and final delivery to the Marshall Space Flight Center, the NASA Center responsible for the ST's development. Perkin-Elmer Corporation was contracted for the design and fabrication of the telescope optics, fine-guidance sensors, and optical support structures, while the Goddard Space Flight Center (GSFC) was responsible for overseeing the development of the Scientific Instruments (SIs), commissioning, and operations.

For the management of the scientific program of this new observatory, a precedent was established; in 1976, the National Academy of Sciences produced a study [4] which recommended that the scientific utilization, optimization, and guidance of the observatory be led by an independent astronomical community-based institution. This *Space Telescope Science Institute* (STScI) was charged with "representing the public interest, as well as the astronomical community and the broader scientific community." This model, though novel and untested at the time, has since become more common practice.

The period from 1977 to the planned launch of the Space Telescope in 1983 was plagued by uncertainties in funding and schedules, many of which show parallels to current James Webb Space Telescope (JWST) experiences. By 1980 it was clear that a launch would not be sooner than 1986. The ST, by now named the Hubble Space Telescope (HST), was originally estimated to cost $600M (USD) to construct. By 1985 the cost had doubled to roughly $1.2B [5]. The loss of Space Shuttle Challenger in 1986 and the subsequent grounding of the shuttle fleet caused major uncertainty as to whether the telescope would ever reach orbit. HST was eventually launched in April 1990, and benefitted from significant progress in ground systems' functionality and readiness during the preceding four years.

The history of HST is a subject that by now has been quite well covered. A detailed account of the HST project history was documented in 1989 in a popular work by historian Robert Smith[6]. A fine early overview of the state of HST and its planned characteristics as of 1982 was given by John Bahcall and Lyman Spitzer [7], while David Leckrone, former HST Project Scientist, provides to a general audience a deft summary from his involved first-hand perspective in [8].

### 3. ORIGINAL SCIENTIFIC GOALS

At the time of HST's launch, we had a very different view of the universe than now. The universe of the 1980s was thought to be decelerating and the expansion rate was greatly uncertain. Black holes at the centers of galaxies were only suspected, and extrasolar planets had not been seen (let alone had their atmospheres' measured). Galaxies were not known to evolve strongly through mergers over time; the notion of hierarchical assembly and structure formation was in its observational infancy. This was the universe HST was released into.

The fundamental scientific questions that could be addressed by a space telescope in general, and HST in particular, were represented at the time by a number of specific "Key Projects" and other areas of study, many of which were planned to be investigated as part of the HST Guaranteed Time Observations (GTO) available to the science teams involved in the development of HST's original SIs, which had capabilities well matched to these pursuits.

These key science goals were established by an independent community-based Space Telescope Advisory Council (STAC). It was expected that these areas would be of greatest interest to the General Observer (GO) and thus comprise a large part of the science program. While this has certainly been true, the periodic enhancements to HST's SIs, the radically different picture of the universe we now have, and the advances in analysis techniques compared with twenty years ago, have meant that many of Hubble's most significant and fruitful areas of investigation to date either were at the time not knowable, not technically achievable, or both. Examples of the notable scientific achievements of the HST mission are given in Section 6.

The three Key Projects established as top priority for HST were:

• Calibrate the cosmological distance scale by determining the Hubble constant, $H_0$, to an accuracy of no worse than 10%. This would be accomplished by using HST to observe "standard candle" objects (in this case a particular class of variable stars) beyond our own galaxy to greatly improve our knowledge of the universe's expansion rate and its age. The accurate determination of $H_0$ would allow tighter constraints to be placed on related parameters like the deceleration parameter $q_0$, and the critical density $\Omega$, all of which pertain to the nature and fate of the universe.

• Determine the properties of the intergalactic medium by observing its absorption signature in the UV spectra of a large number of very distant quasars, caused by the material intervening along the pencil-beam line of sight to each quasar.

• Survey galaxy demographics and other objects of interest by deep imaging of assorted unremarkable regions of the sky. Referred to as the "Medium Deep Survey" (MDS), this project recognized that many of HST's interesting findings might be serendipitous and unplanned.

The Key Projects' observations were "front-loaded" into the early part of the HST mission, and thus were generally hampered by the telescope's famous spherical aberration discussed in Section 4. Some of the data for these programs, however, were taken after Servicing Mission 1 so benefited from the restored optical clarity. These specific early projects have long been completed, and advances in the fields and the observatory's abilities have brought about a plethora of important scientific discoveries and questions, some of which have obviously progressed far beyond what was imagined when these goals were formulated. As two examples, while the cosmological distance scale goal was met[9], subsequent HST observations strongly supporting an accelerating universe have since altered our understanding of the associated cosmological parameters. And the MDS, (hampered by image aberrations and the limitations of a first generation camera), was eclipsed by the Hubble Deep Field and Ultra Deep Field, which both became milestones in observational astronomy.

Other areas that were thought to be well suited to the original HST ran the gamut of astronomical research [10][11]. HST was expected to bring to astronomy a spatial resolution, sensitivity, and observable wavelength range that was unprecedented. As a general purpose observatory, beyond the atmosphere, with initial imaging and spectrophotometric abilities in the optical and UV, the sky was no longer the limit.

## 4. DESIGN AND REQUIREMENTS

SPACECRAFT OVERVIEW

The basic design of HST is depicted in Figure 1. The spacecraft encloses a 2.4 meter (m) diameter telescope (Optical Telescope Assembly, "OTA") and comprises two larger diameter cylindrical sections aft; the "equipment section" ring contains most of the subsystems for power, reaction and logic, while the longer "aft shroud" section sits behind the primary mirror and houses the SIs, rate sensor units (RSU), and attitude observers: the three Fine Guidance Sensors (FGS) and three Fixed Head Star Trackers (FHST).

Attitude control is executed by reaction wheels and sensed by the combination of inertial navigation (RSUs), interferometric nulling (FGSs), and star field matching (FHSTs). Power is provided by a pair of large rotatable solar arrays, which also charge 6 batteries for power during orbital night. Combinations of passive and active thermal control provide careful management of the environment for the optics and instruments.

The telescope diameter of 2.4m was close to the maximum size achievable, being dictated by the need to house the OTA, the spacecraft-related subsystems, and the Science Instruments into a package that would meet the volume and mass constraints of the Space Shuttle program, which was the launch system chosen for its leading payload capacity and unique ability to support the human serviceability requirement on HST. Figure 2 (taken from [12]) illustrates HST's utilization of the Shuttle payload bay.

HST operates in low earth orbit, at altitudes that have varied from 615 km to the current 565 km [13]. During a typical ~96 minute orbit, it will spend most of its time in pointed astronomical observations. This efficiency of ~50% or better was reached during the first 6 years of science operations as improvements were made in how HST was utilized and operated.

Efficiency values predicted at launch were 35%. As a target is observed it will be occulted by the earth anywhere between zero and ~50% of an orbit, at which point HST can halt and resume an exposure after the occultation. Science and engineering data is stored on-board and periodically transmitted to the NASA Tracking Data Relay Satellite System (TDRSS) where it is then sent to the ground.

OTA DETAILS

HST's OTA is a Ritchey-Chrétien Cassegrain telescope design, whose basic configuration and dimensions are shown in Fig. 3, and whose optical parameters are listed in Table 1.

*Table 1: Optical parameters for HST*

**Primary Mirror:**
- 2400 mm diameter circular annulus sandwich design, with 600 mm central opening,
- Corning ultra-low expansion (ULE) glass, $MgF_2$ coating over Al. Hyperboloid: 11040 mm radius of curvature, 5520 mm focal length ($f_1'$)
- Conic constant $K_1$ = -1.0022985 (spec, [14]), -1.0144 (as built, [15])

Note the significant difference between the primary mirror's as-built, and design value for K, which will be discussed later.

**Secondary Mirror:**
- 281 mm diameter (310 mm with housing)
- Schott Zerodur glass, with $MgF_2$ coating over Al. Convex hyperboloid: -1358 mm radius of curvature,
- Conic constant $K_2$ = -1.496 [16]

**System:**
- Focal length (system) $f'$ = 57600 mm, focal ratio (system) = $f/24$, focal ratio (primary mirror) = $f/2.3$, (magnification = 10.43)
- Mirror separation, $t$ = 4906.9 mm, Central obscuration = 33% (diametric)
- Focal surface images ~28 arcmin on sky (~ diameter full moon) at a plate scale from 3.60 - 3.37″/mm
- $MgF_2$ coating thickness is sized to boost UV response but throughput cuts off sharply at ~115 nm.

The Ritchey-Chrétien (RC) design is an attractive choice for modern large Cassegrain telescopes as ideally the mirrors' eccentricities combine to produce no third order spherical aberration or coma. Equations 1 & 2 [17] define the eccentricities of the Primary ($e_1$) and Secondary ($e_2$) in terms of focal lengths and mirror separation such that the RC criterion is met.

$$e_1^2 = 1 + 2f_1'^2(f_1' - t)/tf'^2$$

$$e_2^2 = \left(\frac{f' - f_1'}{f' + f_1'}\right)^2 + \frac{2f'f_1'^3}{t(f' + f_1')^3}$$

*Where approximate values of t, $f_1'$, and f' for HST are given in Table 1, and K = $-e^2$*

The OTA holds HST's secondary and primary mirrors in alignment by a thermally passive graphite-epoxy truss/ring structure (Metering Truss) which effects dimensional control in the thermally dynamic environment. The 4.9 m long truss consists of 48 tubular elements, 2 meters in length, each selected according to its particular measured thermal coefficient of expansion, and then matched to the expected temperature variations for different locations in the truss to minimize overall bending over the structure [18]. The SIs are latched to the Focal Plane Structure Assembly (FPSA) which is mechanically interfaced to the truss via the primary mirror assembly ring (Figure 1). Variations in the dimensions of this truss, over timescales as long as the mission and as short as an HST orbital period (~96 minutes), as inferred from on-orbit science data, are discussed in Section 5. Figure 4 shows the flight Metering Truss and FPSA during construction.

While the truss's temperatures are passively managed, the temperatures of the mirrors themselves require active control to remain sufficiently stable. Section 5 describes on-orbit findings and performance in this area.

Actuation capability is provided by 24 actuators behind the primary mirror. It was expected that these could correct a degree of low spatial frequency figure error if found to be needed. In fact the unintentionally spherically aberrated shape of the primary mirror was beyond the actuators' correction range and they have not been used since the initial commissioning and troubleshooting. The secondary mirror utilizes 6 actuators in the form of 3 bipod supports. This is a common arrangement for providing the full 6 degrees of freedom to an optical element with some redundancy. In practice in routine science operations the bipods have only been used symmetrically to periodically move the secondary mirror in a purely axial motion for focus corrections, also discussed further in Section 5.

Once the 256 nm of known spherical aberration are removed, remaining total system aberrations (OTA+SI) are observed between 25 and 70 nm rms, depending on the SI. The OTA is believed to contribute approximately 25 nm, 18 of which are in the form of mid-frequency "polishing" errors, and the remainder of which is predominantly due to trefoil aberration coming from three primary mirror support pads. See Krist & Burrows [19] for a discussion of phase retrieval techniques used with HST data, the various spherical aberration determinations, and estimates of the wavefront error in the OTA and SIs from 1995.

REQUIREMENTS

Key high level requirements on the HST observatory specified that it be human serviceable by Shuttle crew and that its planned mission life be 15 years. Optically, its Point Spread Function (PSF) would be diffraction-limited at 440 nm, implying ~32 nm rms wavefront error. This implies aberrations at the level of 1/14 wave ($\lambda/14$), 70% of the energy within 0.1 arcseconds (″), and a Strehl ratio (observed/theoretical PSF peak value) of 0.8. Observations at 0.633 μm should show images good to $\lambda/20$, with 80% encircled energy within 0.1″ and a Strehl ratio of 0.9 [20]. Image stability would be good to 0.007″ rms over 24 hours, and the total wavefront error would not degrade to more than another 32 nm rms during that time. See Schroeder[21] for a more detailed discussion of HST's pre-launch expected optical performance.

The OTA would also deliver a flat (within 20%) throughput for the wavelength range of 0.12 to at least 1.1 μm, the range of the original complement of SIs. Providing high quality imaging over a wavelength range that spans from the far UV to ~1 μm (and later out to 2.4 μm) was a challenge that required great attention to the mirrors' micro-roughness, and to their emissivities and overall cleanliness (i.e. minimal areal dust coverage fraction). The latter was particularly difficult to ensure given the multi-year launch slips and the associated storage considerations for the completed coated optics, although evidence discussed later from IR SIs suggests HST exhibits a rather low dust coverage fraction given the time spent on the ground. This will also be an area requiring attention for JWST, the majority of whose primary mirror segments are complete at the time of this writing, an unknown number of years prior to launch.

SCIENCE INSTRUMENTS

The OTA feeds one radial and 4 "axial" SIs (Figure 3), along with three FGSs, which occupy the other radial bays. Unlike many ground-based telescopes where most of the usable focal plane is seen by one SI at a time, HST's ~28 arcminute field of view is spatially shared among the SIs and guiders, each allocated a permanent section of the 0.46m diameter field. Figure 5 illustrates the original SI and guider fields' apportionment within the OTA focal plane.

The initial complement of SIs included two cameras (a CCD-based and a photon counting vidicon tube detector), two spectrographs (high resolution and faint object), a high speed photometer with UV polarimetry, and three FGSs, white-light shearing interferometers, any two of which could be used to control pointing in a 2-Guidestar Fine Lock mode while one was calibrated to also serve as a science instrument, capable of sub-milliarcsecond (mas) resolution.

This original science payload was to detect point sources at Vmag=~30, offer spectral resolutions (λ/Δλ) of 100000 at wavelengths unobservable from the ground, and provide (in a number of modes) critical (Nyquist) spatial sampling of the PSF, with between 2 and 3 pixels across the central 0.1″.

The original complement of SIs flown on HST and their capabilities are well described in e.g. [22][23]. Unique for space telescopes has been HST's periodic replacement and upgrades to its SIs, essentially resulting in a new and more capable observatory each time. See Table 2 for a summary of the twelve SIs, including their acronyms which will be used hereafter. The evolution of the observatory and its performance are covered in the next section.

*Table 2: HST Science Instruments*

**Wide Field & Planetary Camera 1 (WF/PC1) 1990 - 1993 [24]**
CCD Camera. Wide Field Channel: 160″ square, 0.1″/pix   Planetary Channel:  68″ square, 0.043″/pix
Variety of filters, slitless spectroscopy, polarizers
Wavelength range: 0.115 - 1.1 μm
Comments: Radial SI. Replaced with WFPC2

**High Speed Photometer (HSP) 1990-1993 [25]**
Photon-counting tube photometer. Image dissector and photomultiplier tubes. Aperture sizes 1″ - 10″,  Time resolution: >0.01 millisecond
Variety of filters & polarizers
Wavelength range: 0.120 - 0.750 μm
Comments: replaced with COSTAR

**Goddard High Resolution Spectrograph (GHRS) 1990-1997 [26]**
Digicon 1-D diode array. Apertures 0.22″ & 1.74″ (w/COSTAR). Time resolution: >50 milliseconds.
5 1st-order and 1 echelle grating. Spectral resolution (λ/Δλ)= 2000 - 100000 (0.6 - 0.012 Angstroms)
Wavelength range: 0.115 - 0.340 μm
Comments: replaced with NICMOS

**Faint Object Spectrograph (FOS) 1990-1997 [27]**
2 Digicon 1-D diode arrays (red & blue sensitive). Apertures 0.09″ - 3.7″ (w/COSTAR). Time resolution: >30 milliseconds.
7 1st-order gratings and 1 prism. Spectral resolution (λ/Δλ)=200-1300 (5 - 0.9 Angstroms). Spectropolarimetry capability
Wavelength range: 0.115 - 0.850 μm
Comments: replaced with STIS

**Faint Object Camera (FOC) 1990 - 2002 [28]**
Image intensifier + 2-D tube imager. Between 28″ and 7″ FOV, between 0.056″ and 0.014″/pix (w/COSTAR).
Variety of filters, long slit spectroscopy, polarizers
Wavelength range: 0.115 - 0.650 μm
Comments: replaced with ACS

**Corrective Optics Space Telescope Axial Replacement (COSTAR) 1993 - 2009 [29]**
Corrective optical solution in the form of an axial "SI"
3 fixed and 7 deployable reflective elements intercept the aberrated OTA beam to FOS, GHRS, and FOC.
Slightly increased effective focal length for these SIs, and slightly reduced throughput. Dramatically corrected wavefront error.
Comments: replaced HSP

**Wide Field and Planetary Camera 2 (WFPC2) 1993 - 2009 [30]**
CCD Camera. Field split spatially contiguously into four optical channels, seen by separate CCDs. Three ("WF") at 80″ square, 0.1″/pix and 1 ("PC") at 37″, 0.046″/pix.
Variety of filters, including polarizer, and graduated "ramp" filter.
Wavelength range: 0.15 - 1.1 μm
Comments: Corrected OTA aberration with internal optics. Replaced WF/PC1

**Near Infrared Camera and Multi-Object Spectrometer (NICMOS) 1997 - present [31]**
IR Camera & Spectrograph. Three 256 pixel square photodiode arrays (HgCdTe) image separate optical channels: 52″ square @ 0.2″/pix, 19″ square @ 0.075″/pix, and 11″ square @ 0.043″/pix.
Variety of filters, polarizers, and grisms (dispersing elements), slitless spectroscopy. Coronography capability
Wavelength range: 0.80 - 2.5 μm
Comments: HST's 1st cryogenic SI. Extended HST observations out to K band.

**Space Telescope Imaging Spectrograph (STIS) 1997 - present [32]**
Imaging Spectrograph. 2 Multi-Anode Multichannel Arrays (MAMAs) each 25″ square, 0.025″/pix. (NUV & FUV). 1 CCD (VIS) 52″ square at 0.051″/pix.
First-order and echelle spectroscopy through a wide variety of apertures & slits. Imaging, slitless spectroscopy, and coronography ability
Wavelength range: 0.115 - 1.0 μm
Comments: replaced FOS. Was repaired during the last servicing mission.

**Advanced Camera for Surveys (ACS) 2002 - present [33]**
Visible (CCD) and UV (MAMA) imagers. Three channels: WFC (~203″ square @ 0.05″/pix), HRC (~27″ @ ~0.026″/pix), SBC (~33″ @ 0.032″/pix)
Variety of filters, polarizer, grism. Coronography, slitless spectroscopy.
Wavelength range: 0.115 - 1.1 μm
Comments: First large format imager. Replaced FOC. Was repaired during last servicing mission.

**Wide Field Camera 3 (WFC3) 2009 - present [34]**
"Panchromatic imager": UV/VIS (CCD) and NIR (HgCdTe array). UVIS channel (~163″ square @ 0.04″/pix), IR channel (~130″ square @ 0.13″/pix)
Variety of filters, grism spectroscopy.
Wavelength range: 0.2 - 1.7 μm
Comments: Thermo-electrically cooled. Replaces WFPC2. As of installation, is HST's only active IR camera.

**Cosmic Origins Spectrograph (COS) 2009 - present [35]**
UV "point source" spectroscopy through either a clear or neutral density small aperture, 2.5″ diameter. Two channels: NUV (MAMA) 1024 pixels square, and FUV ( cross delay line -type microchannel plate) 32768 x 1024 pixels.
7 gratings for low to medium resolution sprectroscopy (spectral resolution = 2000 - 24000)
Wavelength range: 0.115 - 0.320 μm
Comments: Extremely sensitive. Optical design maximizes system throughput. Replaces COSTAR.

**Fine Guidance Sensor (FGS1R) 1997-present [36]**
Astrometer. Relative astrometry to > 0.2 mas for V<16.8. Detection of binary morphology at 8 mas. 40 Hz relative photometry at millimagnitude precision.
Comments: Serves dual function as a Science Instrument as well as one of the three operational guiders.

# 5. ON-ORBIT EXPERIENCE

PROBLEMS "AS-LAUNCHED"

Upon HST's deployment in April 1990 there ensued a months long period of commissioning and calibrations known as "science verification". During this time of observations and on-orbit tests, which included focus sweeps, secondary mirror tips and tilts, and primary mirror figure control assessments, it was inferred with a high degree of confidence that the HST OTA itself (specifically the primary mirror) was responsible for producing a focal surface with a significant amount of third order spherical aberration (a radially symmetric wavefront aberration), See Figure 6.

In late 1990, NASA commissioned a forensic investigation into the production and testing of the HST primary mirror [37] that identified the cause to be a 1.3 mm spacing error in the reflective null corrector device used to optically test the mirror while being figured. The spacing error resulted from unintended reflections off of a metering rod end cap, rather than from the rod end itself, due to a flaking away of a non-reflective paint from part of the end cap. The calculation of the expected figure error resulting from this hardware and the amount of resulting wavefront aberration agreed well with later measurement of stellar PSFs using phase retrieval techniques, with values for the primary mirror's conic constant K varying from -1.0132 [38] to -1.0144 [39], compared with the design value of -1.0023 necessary to meet the Ritchey-Chrétien condition. Corresponding rms wavefront error estimates range from 253 [40] to 277 nm [41]. Surface error at the edge of the primary mirror is approximately 2.3 μm. Subsequent corrective optics have not all been built to the same prescription, but all have assumed a value within this range, and the differences have not been significant. Krist & Burrows [42] discuss the refinement of HST's prescription and quantify residual errors after the first generation of corrective instruments was installed during the first HST Servicing Mission.

A committee which in 1990 studied a number of strategies for addressing the spherical aberration produced a report [43] describing the problem and the proposed solutions.

In addition to its optical problem, HST's two large flexible solar arrays were causing the Pointing Control System (PCS) to perform outside of specifications. The original solar array design consisted of two flexible rectangular wings, each spanning 12 x 2.8m, for a total collecting area of ~70m$^2$. Together they produced 4kW at 34V. The very thin substrate on which the cells reside was wound about a cassette at the end of a deployable arm, and was pulled out into the fully extended operational position by a thin metallic bi-stem arrangement. Flimsy by design, these large arrays and their bi-stem supports were expected to smoothly deform in response to the extreme temperature swings as HST made terminator crossings between orbital day and night. On-orbit however, it was found that one of the arrays would abruptly "snap" into a new position as it thermally deformed. This impulse was often sufficient to cause the FGSs to lose interferometric lock on guidestars, usually resulting in a loss of science.

To mitigate this, the FGSs were run much more frequently in a more forgiving "coarse track" mode, but this prematurely wore the bearings of the guiders' mechanical star selector mechanism as it nutated about the guidestar's center of light. A fight software modification to the PCS was then developed which changed parameters in the pointing control law and the FGSs. This had the effect of "stiffening" the control loop and protecting the FGSs from the loss of lock associated with the transients, but was costly in terms of on-board resources. The first servicing mission replaced these solar arrays with a slightly modified design (Figure 7), and in 2002, a third pair of rigid folding and more efficient solar arrays was installed. In addition to having physical disturbances with a power spectrum more benign to the PCS, the smaller arrays made 5.7 kW of power, supporting the increased power demands of the 2nd and 3rd generation SIs.

Other problems were manifest between the time of HST's deployment and the first servicing mission, among them premature failures in multiple RSUs and impaired performance of the Magnetic Sensing System (MSS). During servicing, these units were replaced, and upgrades to an on-board computer (DF-224), and an SI (GHRS) were successfully performed.

The potential of HST, however, did not go unnoticed when it became clear that, despite these problems, and the fact that the observatory did not meet its highest level pointing stability and image quality goals, the early science results were significant, and already exceeding what was possible from the ground [44]. Especially good success was being obtained with image deconvolution techniques, which were being quickly adapted and enlisted to deconvolve the spherically aberrated (but known) PSF from the astronomical sources in HST images [45][46]. See Figure 8.

SERVICING MISSIONS & EXPANDED CAPABILITIES

Since 1993 five manned servicing missions to HST have replaced SIs with more modern incarnations, and maintained and enhanced the observatory. Figure 9 graphically summarizes the missions, highlighting the evolving focal plane, while Table 2 outlines the basic characteristics of the complement of SIs, past and present.

Information on the HST servicing missions is widely available, for example from NASA [47] and ESA [48] websites. A recounting of the high drama and technological challenges associated with human servicing of HST is beyond the scope of this paper, but the following significant milestones in HST's servicing, most relevant to its history, are (admittedly subjectively) called out.

Servicing Mission 1 (SM1, Dec. 1993) exchanged the HSP for the Corrective Optics Space Telescope Axial Replacement (COSTAR), which deployed optics intercepting the OTA's beam, providing a spherical-aberration-corrected feed to the existing first generation SIs[49]: GHRS, FOS, and FOC. The WF/PC1 camera was replaced by WFPC2, which featured its own internal corrective optics. Together with the improved solar arrays, these replacements allowed HST to realize its intended performance. See example in figure 10. After correction the system provided diffraction-limited imaging, with $\lambda/21$ waves rms wavefront error @ $\lambda = 0.5$ μm, with Strehl ratios of ~90% and PSF core widths of 50 mas at $\lambda = 0.55$ μm [50].

HST images were well underway to becoming globally recognized icons, and it is amusing to note the persistence among the public, the media, and even educators, of the "lens" or "eyeglass" concept for HST optical correction, when in fact reflective elements have been used in all cases but one; WFC3 IR channel utilizes a correcting lens for packaging reasons.

Servicing Mission 2 (SM2, Feb. 1997) is perhaps most notable for the installation of what was HST's first cryogenic SI, extending the observatory's potential into the IR, 2.5 μm. NICMOS probed the performance of the OTA in this long wavelength regime where mirror temperatures and emissivities are critical. In order to achieve required mirror figure stability, HST's primary and secondary mirrors are heated, with temperatures controlled around a setpoint of 21.1°C [51]. Due to gradients across the mirror from front to back, actual temperatures measured near the primary mirror surface on the outer and inner edges indicate temperatures between 14° and 16°C that are stable to 1°C. The secondary mirror surface is similarly stable and cooler at the surface than the control point. The cooler temperatures have benefitted HST's IR performance. At the same time, the temperatures are sufficiently stable for a constant mirror figure whose changes are not clearly observable in aberration monitoring so are very small compared with the varying focus aberration discussed later. Details of HST's performance as an IR telescope and the measured or inferred temperatures and emissivities are given by Robberto et al. [52].

NICMOS's dewar contained nitrogen ice, whose pre-launch dimensional changes deformed a housing sufficiently to cause thermal contact with a surrounding shield once in orbit. This resulted in the premature depletion of the cryogen and warming of the instrument by early 1999. Cryogenic temperatures were again reached via a cryocooler[53] installed in 2002, restoring NICMOS's scientific productivity for another six years. Since late 2008 the cryocooler has failed to run consistently, effectively suspending NICMOS science for the time being.

The intended Servicing Mission 3 was split into two separate missions just over two years apart. SM3A (Dec. 1999) was flown expeditiously to replace the 3 RSUs (containing 2 gyros each) used on-board for attitude sensing and pointing feedback. Two RSUs had already been replaced during SM1 but by 1999 half of the gyros had failed for reasons that were largely understood. Six weeks prior to the SM3A, a fourth gyro failed. At the time HST required a minimum of three gyros to operate and thus was put into a "safe-mode" attitude. For the first and only time to date, HST was performing no science for an extended period (~40 days). New gyros, an upgraded spacecraft computer (the original was '60s-era technology), and a solid-state digital data recorder (the originals were reel-to-reel models) were notable among a host of other refurbishments.

Servicing Mission 3B (SM3B, Mar. 2002) was a milestone for the installation of the ACS, a powerful modern imager whose "discovery power" (the product of sensitivity and field of view) was many times greater than the existing cameras. ACS quickly became responsible for some of HST's most notable scientific accomplishments and iconic imagery, but in early 2007 after 5 years of productivity it suffered a major failure in parts of its power circuitry. Restoration of a significant portion of its science capabilities would take place in SM4.

SM3B also featured the highly challenging but successful work to install the NICMOS cryocooler with its circulating fluids, and the externally mounted large radiator associated with the Aft Shroud Cooling System. This mission also included tedious fine-scale work to replace HST's Power Control Unit (PCU). Until SM4, SM3B had served as the best example of the complex tasks humans can accomplish in space.

Servicing Mission 4 (SM4, May 2009) was originally scheduled for 2004, but after the loss of the shuttle Columbia it was canceled. A later reassessment of the decision resulted in reinstating the mission, which was scheduled for October 2008. With servicing years overdue, HST's SIs, gyros, and other systems were degrading and failing, and in late September 2008, a mere three weeks before SM4 was scheduled to fly, HST's SI Control and Data Handling (SIC&DH) computer failed and the backup was now operating with no redundancy. Not wishing to fly a final HST servicing without addressing a critical failed component of the observatory, NASA postponed the mission while a replacement computer was readied and the crew trained for its installation.

By May 2009, SM4 was underway, with a daunting manifest. It saw the successful installation of WFC3 and COS, both representing major leaps in HST's scientific utility (see Figure 11). It restored the Wide Field channel to ACS, and resurrected the highly capable STIS, which was installed in 1997 (SM2) and failed due to power circuitry problems in 2004. Aside from installing or repairing a total of four SIs, the new SIC&DH was supplied, all six gyros were replaced, a new FGS was installed, the original spacecraft batteries were replaced, new thermal insulation was laid, and a capture mechanism for potential future use was connected.

Working physically within the SIs as was done during SM4 was not a mode for servicing HST that was envisioned when the mission was designed, nor was it planned for when these SIs themselves were designed and built years later. The ACS repair was described as the most difficult work performed during a space walk[54]. The public statement from NASA headquarters on SM4[55] asserts that *"this repair program represented a significant step forward in the ability to perform on-orbit spacecraft servicing."*

PERFORMANCE & BEHAVIOR

At over 20 years on-orbit, HST's performance and behavior of its numerous complex subsystems cannot be covered even in summary form in anything less than volumes. HST is important for what it teaches us about space observatories, as well as the universe. The very few areas described below are those which 1.) are systems-level and broad, 2.) involve the scientific productivity of the observatory, and 3.) say something about the interaction of the spacecraft with its environment, and thus of general importance to space telescopes.

Optics:

To first order, the HST OTA provides a stable PSF to the SIs: it obviously has no scintillation. It does not vary due to mirror deformations, and there is also no evidence that the metering truss fixing the secondary and primary mirrors introduces measurable non-axial element motions such as tip, tilt, or decenter into the system, at least not at the level to produce detectable aberrations or optical axis misalignments. However, the telescope does experience noticeable changes in focal length as the length of the truss (or parts thereof) and hence the axial separation between the mirrors, varies over timescales as short as an orbit, and as long as the mission life[56].

This change in the physical position of the secondary mirror along the OTA axis of symmetry ("despace") induces observable changes in focus as illustrated by Figure 12, which shows a typical variation over an HST orbit, as measured using phase retrieval techniques on ACS data and as modeled using a temperature-based focus model which has evolved over time [57][58][59]. 1 micron (μm) of despace at the HST secondary mirror induces 6.1 nanometers (nm) of rms wavefront error [60]. We observe orbital swings of typically +/-3 μm, though the amplitude varies and up to ~7 μm of despace have occasionally been experienced. Corresponding focus changes at the wavefront would be +/- 18 nm ($\lambda/28$ at 0.500 μm) and up to $\lambda/11$ in rare circumstances. Characterizing and addressing this focus variability's specific effects on the PSFs at a given SI is important to the most accurate analyses of particularly sensitive science data, e.g. [61][62][63].

In addition to the secondary mirror despace oscillating with an orbital period, it also wanders irregularly over timescales of many orbits to months by an amount generally comparable to the peak to peak orbital changes. The temperature changes responsible for inducing these secondary mirror despaces are driven by various attitude-related parameters associated with HST in low earth orbit. Figure 13, taken from [64] illustrates the influences of these parameters on focus. Though an attitude-based focus model, as accurate as the temperature-based model, has not been achieved, some of the main drivers have nonetheless been identified.

HST also exhibits a secular, long term shrinkage of its graphite epoxy truss that has persisted far longer than expected pre-launch. The truss was expected to shrink in a roughly exponential manner as moisture in its material desorbs (migrates out and into space). It was thought that total shrinkage would effectively stabilize after 3 to 4 years on orbit.[65] In fact, from long term phase retrieval measurements of HST PSFs we have observed a shrinkage of over 0.150 mm, which is still occurring today at a level that requires periodic active focus adjustment. See Figure 14. The curve is fit well by the following double exponential

$$\Delta SM = -5.04 + 56.26 e^{-t/364.53} + 106.24 e^{-t/2237.2}$$

*where $\Delta SM$ is secondary mirror despace in microns and t = mission elapsed days.*

Note that the short term exponential does indeed decay after a few years, while the unexpected term's longer e-folding time may suggest desorption of other types of molecules or possibly other processes at work. We are not aware of a graphite epoxy truss, with telemetered or inferred metrology, in space vacuum for the duration of HST, so our on-orbit experience (monitored from the perspective of maintaining image quality) may be of potential interest in other disciplines.

This shrinkage over 20 years totals only $3 \times 10^{-5}$ the length of the truss, yet represents nearly 2 waves of defocus and has resulted in the commanding of over 20 compensatory secondary mirror despace adjustments away from the primary mirror to maintain the desired high level of image quality.

When new SIs are installed on HST, part of their commissioning and internal optical alignment involves achieving the best possible degree of confocality with existing SIs. See Hartig[66][67] for recent examples. Image measurements have indicated that confocality can typically be achieved to within 6 nm rms wavefront error (~20% of typical orbital focus oscillations). HST focus maintenance is thereafter executed via the OTA secondary mirror moves rather than internal optical adjustments in each SI.

Pointing:

After the optical requirements and performance, the pointing is another critical parameter for a space telescope. Here again, HST has performed very well throughout its mission, exceeding requirements, but exhibiting similarly interesting behavior on varying timescales. Presented here is experience with HST's absolute accuracy, and its pointing stability.

Absolute astrometry, like any other absolute calibration, is fundamentally a bootstrapping incremental process. Over the long HST mission life, the astronomical community's catalogs and surveys have converged towards the current common reference frame, the International Celestial Reference System (ICRS)[68] and significant systematic differences between catalogs have become less common. Over the same period advances in observational methods, computing power, and data reduction methods have reduced relative position errors within catalogs considerably.

When the HST mission began, it required a new star catalog, created to support guide star selection. The Guide Star Catalog 1 (GSC1) [69] was the largest catalog of its time, containing 19 million entries, with magnitudes down to ~15. HST in its most accurate pointing mode holds two guidestars in an interferometric lock somewhere along the perimeter of its field of view. The accuracy of Hubble's PCS is intrinsically much better than the two biggest contributors to real-world pointing error: 1.) the error in the guidestars' known catalog positions and 2.) the alignment errors between the SIs and FGSs, and to a lesser degree the distortion errors within them, both of which are time dependent. Figure 15 (adapted from [70]) is an illustrative example of the positional change over time of a newly flown instrument, in this case an FGS. This is believed to be due to physical motions within the instrument as it it desorbs moisture in a manner similar to the OTA's truss. Increasing position error affects the absolute pointing and astrometry during the times between the periodic ground system calibration updates to characterize the location.

GSC1 (utilized as the default guidestar catalog for HST until 2006) contained 1 sigma errors of 0.7″ relative within the catalog. It also exhibited a significant systematic with respect to ICRS, and earlier in the mission offsets of up to 2″-4″ could be seen between a given object's positions in GSC1 and an ICRS-based catalog like Hipparcos. With these errors and position evolution in the focal plane, 1 sigma pointing accuracy of ~1″ (or worse if target coordinates were specified in an ICRS-based catalog) was not uncommon.

Guide Star Catalog 2 (GSC2) [71] began use in HST operations in 2007. GSC2's relative errors were between 0.15″ - 0.2″, and the catalog was now referenced to ICRS. Since 2007, the use of GSC2 along with more frequent calibration of the HST focal plane model to below the catalog error level has resulted in absolute pointing accuracy of HST of ~ 0.2″ 1 sigma [72]. It has also been observed that COS and WFC3 have been more dimensionally stable than past orbital replacements, which has aided this effort.

HST's pointing *stability* is by most definitions excellent, but as with the focus stability, at low levels and for sensitive science, it is measurable and dynamic over multiple timescales, and must be taken into account. While the high frequency "jitter" is excellent, at the level of < 5 mas 1 sigma, and does not measurably smear the PSF (in non-pathological cases), the pointing is seen to trend over various timescales. Recall the 0.007″ stability expectation over a timescale of orbits. There are significant differences in the effect and handling of such a value if it is incurred as high frequency random jitter, or if a smooth, long-term position drift. On the JWST project, the appropriate specification of the various types and timescales of line of sight changes is receiving particular attention at the time of this writing.

As mentioned earlier, in most cases, HST's science target is not located in the instantaneous continuous viewing zone (CVZ) so is occulted by the earth every orbit. If observed over multiple orbits, the reemergence of the target will be followed by a guidestar "re-acquisition", in which the FGSs will search for and usually acquire the guidestar pair close to their last known location, though having drifted due to normal gyro error accumulated during the occultation. While observing multiple exposures within an orbit, multiple precise measurements at the SIs generally will reveal smooth drifts of between 5-10 mas. After a re-acquisition of the same guidestars a discontinuous jump of a comparable amount is commonly observed, and the smooth drift begins again. On top of this sawtooth effect associated with observing over multiple orbits there is also a long-term decaying drift seen when a target is observed continuously over timescales of days. This effect has been seen to produce between 25 - 50 mas of drift, with most of that occurring within the first day. See Figure 16, adapted from [73]. The hourly/orbital changes equates to 0.1- 0.2 pixels for a 50 mas pixel e.g. ACS/WFC, while the range seen in the long term settling over days is at the ~1 pixel level. This basic behavior has been observed with multiple imagers over the mission life, although specific SIs, and specific thermal influences create considerable variations in the magnitude and overall behavior. See Gilliland 2005[74] for a quantitative case-study of these effects as a by-product of his data reduction of science observations involving long-duration (multi-day) pointings.

From August 2005 until science resumed after SM4 in 2009, in order to conserve gyros while awaiting the overdue servicing, HST science operations were performed with 2 gyros in the control loop rather than the normally requisite three. On-orbit testing of this mode as well as statistics from ensuing science showed that jitter and drift during observations taken under normal 2-guidestar fine guiding showed no signs of degradation. In fact jitter characteristics and pointing performance were in practice indistinguishable[75]. The reduced gyro mode did however affect operations by decreasing slightly the observatory efficiency, as more time was spent on PCS-related setup and attitude determination between guidestar acquisitions. In addition, the incomplete inertial attitude sensing increased the chances of a failure to acquire and/or achieve interferometric lock of guidestars, a problem that was still rare. The success of 2-gyro mode for HST was encouraging and has led to the specification of a viable 1-gyro mode for potential future implementation.

While the six gyros within the three RSUs provide the rate sensing, four 45kg reaction wheel assemblies (RWAs) provide the control torques for maneuvering (3-axis rotation); there are no propellants on-board HST, which cannot actively translate itself. In SM2, one of the RWAs was replaced and since that time all four have been functioning nominally. Options are being investigated however, for science operations using a reduced number of reaction wheels; three- and even two-RWA modes are being considered, should one or more fail in the future.

The main experiences with the stability of HST's focus and pointing has shown that the system is too complex to allow the practical development of predictive or descriptive models that are sufficiently complete and accurate. An important point in this regard is that the HST science efficiency is placed at the highest priority, and the inordinate effort and large amounts of dedicated HST observing time required to explore the parameter spaces necessary to better support such model development are prohibitive. Rather, the successful science programs whose results are sensitive to such instrumental effects usually involve empirical corrections or accounting of these effects, for example see [76][77]. As the mission has evolved and altered our understanding of the universe, HST has been used for ever more challenging observations, "pushing the limit" of what is possible with the observatory, and exceeding what was imagined when it was conceived.

Contamination:

An important lesson in space-borne optics was learned by the HST mission "the hard way" regarding the risk of contamination affecting UV-sensitive optics. Initial deployment and regular visitations to HST on-orbit have brought to the telescope not only a biological presence, but hardware coming from atmosphere. As pointed out earlier, in normal science operations, HST often observes targets across earth occultations, suspending and resuming exposures accordingly; there are normally no earth avoidance restrictions for its line of sight. After the original WF/PC1 camera was replaced and returned to the ground after SM1, its pick-off mirror, fully exposed ahead of the axial SI enclosures, was found in laboratory analyses to be heavily contaminated with organic polymers and showed very low throughput in the far UV. Molecular contaminants, outgassed and desorbed from the observatory following deployment, and photo-polymerized by the bright earth as it regularly passed across the line of sight, was understood to be the cause. From SM2 onward, the Servicing Mission Observatory Verification (SMOV) period of calibrations and commissioning included a "bright earth avoidance" period combined with UV sensitivity monitoring [78], and thereafter, active mitigation procedures. The intent was to avoid photo-polymerization of the outgassing products, which are released at an initially high rate from newly installed hardware.

This is an early example of contamination on-board HST, and areas such as the nature, source, and optical response of contaminants, their deposition as a function of temperature, and effects of periodic warming or other active mitigation remain important topics for understanding, especially when UV performance must be well understood and characterized for science. The COS instrument's UV sensitivity is currently being trended and characterized, and attempts are being made to understand the cause of an observed decline due to a possible contaminant or other environmental causes.[79]

Radiation damage - detectors:

One of the more significant phenomena experienced over HST's long duration on-orbit is the pervasive radiation damage to CCD detectors[80]. High energy particles cause defects in a CCD, many of which "trap" charge as it is transferred out, especially in the parallel transfer direction. These traps, which increase in number with continued exposure, reduce the detector's charge transfer efficiency (CTE). Many of these traps then release their charge at some later time. The clocking timescales, the number of pixels over which the charge is transferred, and the trap release times, together have the effect in the science data of removing flux from the core of the PSF and distributing it asymmetrically in a linear "tail" along the parallel direction. This can hinder science that involves photometry (by removing flux from a given central radius), astrometry (by affecting the centroid due to asymmetry), and morphology. This effect steadily worsens over time (exposure), and in practice is unavoidable. It can though be mitigated through a variety of modeling and calibration methods, e.g. [81], and significant further progress has been made in these areas recently[82][83].

Radiation damage - exterior:

Vital to HST's passive thermal control in low earth orbit is the extensive use of low emissivity Multi-layer Insulation (MLI) to wrap the majority of spacecraft's exterior, which sees ~250°C orbital temperature swings. MLI consists of 15 layers of aluminized Kapton, with an outer layer of aluminized Teflon. After SM1 and more noticeably after SM2, this material was found to be eroded, with visible damage (cracking and peeling) especially to the outer teflon layer in a number of places. HST science data and engineering telemetry indicated the degraded MLI was not at that time causing changes in the observatory's thermal state. A review board[84] in 1998 analyzed MLI samples returned during servicing, determined the cause (particle and UV/Xray flux, combined with thermal cycling), and recommended a number of similar but more robust materials and constructions to be flown in subsequent servicings, replacing damaged MLI in particularly sensitive areas, to protect against worsening of the thermal stability over time. Since then, these improved MLI patches, and other specialized replacement fabrics and foils have been flown and placed over existing materials in various locations at various times in all subsequent servicings, and it was encouraging to find during SM4 that the overall condition of the MLI was better than had been predicted.

ADVANCES IN OBSERVING, ANALYSIS, AND ARCHIVING

Major advancements have been made in the way HST is used, in the techniques to analyze data, and in processing, distribution, and availability of high level science products.

Not long after HST began its science mission, the development and implementation of "parallel" observations took advantage of the spatial sharing of the HST focal plane among the SIs by allowing multiple instruments to image the sky at once, a mode which has become even more attractive and popular given the current suite of SIs. The concept of "dithering" (small pointing offsets at multiple pixel and sub-pixel levels) with HST imagers became routine, and useful to take advantage of new data reduction techniques like "drizzle"[85] that were being developed to combine, clean, and resample images.

In the post-observation ground system, the model of the science data "pipeline" to produce routinely calibrated science products was an integral part of the HST observatory operations from the onset. Since then, On the Fly Re-processing (OTFR) was implemented mid-mission, whereby data requested from the archive would automatically be re-reduced using the latest calibration knowledge applicable to the time of the observation.

Better calibration tools and new processing techniques have continually evolved over the HST mission. Two recent examples are the CTE correction method discussed in [86][87] and available for use at [88], and the tool "Multidrizzle" (described and made available at [89]).

To investigate key scientific questions and disseminate high quality results to a wide audience, HST time is being increasingly used for long-term or high volume science programs, usually run by large teams that feed back to the community and the HST archives "value-added" high level science products.

Repositories of highly processed science data, often featuring accuracies improved at a later time beyond their original levels, are growing in popularity and utility (as well as in volume). The Hubble Legacy Archive (HLA)[90], and the Virtual Astronomical Observatory (VAO)[91] are examples of such resources. These "high-level" data might be for example a large mosaic of tens of arcminutes. This product would have been built by combining numerous separate HST observations, at different rolls, using different guidestars, and even taken during different years in a way that retains or improves the data's fidelity. Dramatic advances in computing and storage over the 20 years of the mission have been key to such developments, and the future is probably well represented by a model of intuitive natural interfaces to vast repositories of consolidated, multi-mission, and even multi-disciplinary high level science data sets that can be continually dynamically improved and readily accessed by all.

In a fictitious illustrative example, a modern science program might observe a wide area of the sky (e.g. ~10 arc-minutes square) with two large format imagers: ACS, and WFC3/IR. The savvy GO would likely carefully choose a dithering strategy benefitting both SIs, but combine this with a larger-scale mosaicking on order of arcminutes. CTE corrections would be applied during data reduction. Multidrizzle would clean, geometrically flatten, and resample the data from both cameras. Astrometric metadata could be improved over the default by fitting multiple star coordinates to their geometrically corrected pixel locations, and a final product would be a deep and expansive (10 arcminute), high-resolution, multispectral swath of sky covering from the IR to the UV, with accurate astrometry and photometry, supplied to the multi-mission high level archives and available to the community as a rich asset to be studied for years into the future.

The STScI has held a number of workshops between 1993-2010, gathering and presenting a wide range of methods of data reduction, new characterizations or modeling of the telescope or SIs, observing methods, archive advances, and related work. The proceedings and websites associated with these "Calibration Workshops"[92][93][94][95][96] capture some of the advances over the years in how HST is utilized and optimized.

# 6. IMPACT OF HST

During this writing, HST made its millionth observation. The significant findings made or supported by HST observations, often in conjunction with other missions on the ground and in space, are numerous and well documented not only in the technical literature, but widely throughout the mainstream media.

Referring back to the original key projects in Section 3, we find that since SM4, observations with WFC3 have allowed the Hubble constant $H_0$ to be determined to 3%. When taken in conjunction with data from other missions, this has tightened the constraints on the nature of dark energy and even suggests an undiscovered species of neutrino. [97]

Regarding the intergalactic medium (IGM) and its baryonic content, the COS instrument, installed in SM4, was designed with this area in mind, and is in the process of taking on this study with an efficiency and capability that was not imagined when it was proposed as a key project. The increased sensitivity of COS allows 50 - 100 times more quasars and thus lines of sight into the IGM to be probed.

Similarly, the Medium Deep Survey key project, only moderately fruitful at the time, saw its intent realized in spectacular fashion by the Hubble Ultra Deep Field (HUDF) which exposed on an "empty" area of sky for 412 HST orbits, providing a wealth of data and surprises which led to numerous findings, collaborations, follow-on research, and coordinated multi-mission follow-up. Additionally, it provided some of the most compelling and popular HST images, evoking a spray of gemstones, each one a young galaxy. Following this successful model for doing fundamental science, today, HST is undertaking the three largest collaborative programs to date, 500-1000 orbits spread over 3 years as a part of its science program. These will be probing the Andromeda galaxy, performing an extensive sky survey, and imaging distant galaxy clusters.

Hubble's observations of very distant supernovae have been key to establishing an accelerating universe, implying the existence of "dark energy", one of the most fundamental revelations of our time. Closer to home, HST has been a critical tool in understanding the details of extrasolar planets. It made the first direct visible light image of such a planet, and observed its orbital motion. It has measured the chemical composition of the atmospheres of others. Extremely challenging observations, pushing the limit of our ability to calibrate HST, are currently probing the distribution of dark matter by the minuscule distortion that its gravity has on the shape of everything we observe.

With its new camera, WFC3, HST has just recently detected the earliest, most distant galaxies known. At redshifts between 8 -10, they show us the 13.8 billion year-old universe as it was just 5 hundred million years after the big bang. NASA Adminstrator Charles Bolden was the pilot of the shuttle mission that carried HST to orbit. In a press release he commented on these observations saying, *"We could only dream when we launched Hubble more than 20 years ago that it would have the ability to make these types of ground-breaking discoveries and rewrite textbooks."*[98]

Metrics used to gauge the productivity and impact of an observatory illustrate HST's influence. A traditional record of research and discovery is the publication of results in peer-reviewed journals; in late 2009, just prior to SM4, there were ~8,300 refereed papers based on HST data and ~287,000 citations. The papers themselves show an observatory's productivity, while the citations indicate its *impact*, and HST has generated on average over 1 published paper in a major journal, and 37 citations to it, every day over its 21 year mission to date.

HST is currently at the height of its scientific capabilities. With no future servicing planned or even practical, STScI and GSFC have been developing ways of retaining the observatory's performance as far as possible into the future. The past successful experience with 2-gyro mode has led to the development of a 1-gyro mode. Operations that use a reduced number of reaction wheels are also being studied, as will be other life extension plans.

There will come a time, however, when HST is no longer able to do science. The JWST, its most obvious successor, is currently facing budget uncertainties not unlike those that troubled HST on its tortuous path to realization, those which seem intrinsic to any singular endeavor that pushes the boundaries of what is possible today, to reveal to us the unknown tomorrow. In 2011, as Hubble presses on past its millionth exposure, and the U.S. weighs the cost of its successor, we are reminded of Lyman Spitzer, who in 1979[99], said: *"The uncertainties of the space astronomy program may remind us of the quotation from T.S. Eliot's poem,* The Hollow Men:

*Between the idea and the reality*

*Between the motion and the act*

*Falls the shadow"*

ACKNOWLEDGMENTS

The author wishes to thank George Hartig, Ken Sembach, Olivia Lupie, John Krist, Colin Cox, Stefano Casertano, Helmut Jenkner, Roeland van der Marel, and Carl Biagetti, and recall deeply missed colleagues and friends Russ Makidon and Rodger Doxsey for their immeasurable impact.

# REFERENCES

[1] L. Spitzer Jr., "Report to Project Rand: Astronomical Advantages of an Extra-Terrestrial Observatory (1 September 1946)," *Astronomy Quarterly* **7**, 131 (1990).

[2] L. Spitzer Jr., *Scientific Uses of the Large Space Telescope,* Ad Hoc Committee on the Large Space Telescope, Space Science Board, National Academy of Sciences-National Research Council, Washington, DC (1969).

[3] L. Spitzer Jr., "History of the Space Telescope," *Quarterly Journal of the RAS* **20,** 29-36 (1979).

[4] D. Hornig, Ed., *Institutional Arrangements for the Space Telescope*, *Report of a Study at Woods Hole, Massachusetts, July 19-30, 1976,* Space Science Board, National Academy of Sciences-National Research Council, Washington, DC (1976).

[5] J. K. Beatty, "HST: Astronomy's Greatest Gambit," *Sky & Telescope* **69**(5), 409-414 (1985).

[6] R. W. Smith, *The Space Telescope: A Study of NASA, Science, Technology and Politics*, Cambridge University Press, New York (1989).

[7] J. N. Bahcall & L. Spitzer Jr., "The Space Telescope," *Scientific American* **247**(1) 40-51 (1982).

[8] D.S. Leckrone, "Hubble Space Telescope," in *Encyclopedia of Space Science and Technology*, H. Mark, Ed., p.760, Wiley & Sons, New York, NY (2003).

[9] W. L. Freedman, B. F. Madore, B. K. Gibson, L. Ferrarese, D. D. Kelson, S. Sakai, J. R. Mould, R. C. Kennicutt, Jr., H. C. Ford, J. A. Graham, J. P. Huchra, S. M. G. Hughes, G. D. Illingworth, L. M. Macri, P. B. Stetson "Final Results from the Hubble Space Telescope Key Project to Measure the Hubble Constant," *Astrophysical Journal* **553**(1), 47–72 (2001).

[10] E. J. Chaisson & R. Villard, *The Science Mission of the Hubble Space Telescope*, Space Telescope Science Institute preprint series #421, pp. 34-41, Baltimore, MD (1990).

[11] M.S. Longair & J.W. Warner, Eds., *Scientific Research with the Space Telescope*, *IAU Colloquium No. 54*, U.S. Government Printing Office, Washington, DC (1979).

[12] R. Carter, *Space Telescope Systems Description Handbook*, *ST/SE-02*, p.274, Lockheed Missiles & Space, Company, Inc., Sunnyvale, CA (1985).

[13] R. Strafella, private communication (2008).

[14] L. Allen, R. Angel, J. D. Mangus, G. A. Rodney, R. R. Shannon, C. P. Spoelhof, *The Hubble Space Telescope Optical System Failure Report*. NASA Pub. TM-103443, p.E-2, NASA, Washington, DC (1990).

[15] J. E. Krist, C. J. Burrows, "Phase Retrieval analysis of pre- and post-repair Hubble Space Telescope images," *Applied Optics* **34**(22), 4951–4964 (1995).

[16] *Ibid*. 14

[17] V. N. Mahajan, *Optical Imaging and Aberrations, Part 1*, p.416, SPIE Press, Bellingham WA (1998).

[18] R. Carter, *Space Telescope Systems Description Handbook*, *ST/SE-02*, p.126, Lockheed Missiles & Space, Company, Inc., Sunnyvale, CA (1985).

[19] *Ibid*. 15

[20] C. J. Burrows, J.A. Holtzman, S. M. Faber, P. E. Bély, H. Hasan, C. R. Lynds, D. Schroeder, "The Imaging Performance of the Hubble Space Telescope," *Astrophysical Journal* **369**, L21-L25, (1991)

[21] D. J. Schroeder, *Astronomical Optics*, p.213, Academic Press, San Diego, CA (1987)

[22] J. K. Beatty, "HST: Astronomy's Greatest Gambit," *Sky & Telescope* **69**(5), 410-412 (1985).

[23] E. J. Chaisson & R. Villard, *The Science Mission of the Hubble Space Telescope*, Space Telescope Science Institute preprint series #421, pp. 8-17, Baltimore, MD (1990).

[24] J. MacKenty, R. Griffiths, W. Sparks, K. Horne, R. Gilmozzi, S. Ewald, C. Ritchie, S. Baggett, L. Walter, and G. Schneider, *WF/PC-1 Instrument Handbook, version 3.0*, Space Telescope Science Institute, Baltimore, MD (1992).

[25] R. C. Bless, J. W. Percival, *High Speed Photometer Instrument Handbook, version 3.0*, Space Telescope Science Institute, Baltimore, MD (1992).

[26] D.R. Soderblom, A. Gonnella, S.J. Hulbert, C. Leitherer, A. Schultz, and L.E. Sherbert, *GHRS Instrument Handbook, version 6.0*, Space Telescope Science Institute, Baltimore, MD (1995).

[27] C. D. Keyes, A. P. Koratkar, M. Dahlem, J. Hayes, J. Christensen, S. Martin, *Faint Object Spectrograph Instrument Handbook, version 6.0*, Space Telescope Science Institute, Baltimore, MD (1995).

[28] A. Nota, R. Jedrzejewski, M. Voit, W. Hack, *Faint Object Camera Instrument Handbook, version 7.0*, Space Telescope Science Institute, Baltimore, MD (1996).

[29] M. Bottema, "Reflective correctors for the Hubble Space Telescope axial instruments," *Applied Optics* **32**(10), 1768–1774 (1993).

[30] M. McMaster, J. Biretta, Eds., *Wide Field and Planetary Camera 2 Instrument Handbook, version 10.0,* Space Telescope Science Institute, Baltimore, MD (2008).

[31] A. Viana, T. Wiklind, A, Koekemoer, D. Thatte, T. Dahlen, E. Barker, R. de Jong, N. Pirzkal, *Near Infrared Camera and Multi-Object Spectrometer Instrument Handbook for Cycle 17,* Space Telescope Science Institute, Baltimore, MD (2009).

[32] K. A. Bostroem, Ed., *Space Telescope Imaging Spectrograph Instrument Handbook for Cycle 19*, Space Telescope Science Institute, Baltimore, MD (2010).

[33] A. Maybhate & A. Armstrong, Eds., *Advanced Camera for Surveys Instrument Handbook for Cycle 19*, Space Telescope Science Institute, Baltimore, MD (2010).

[34] L. Dressel, Ed., *Wide Field Camera 3 Instrument Handbook for Cycle 19*, Space Telescope Science Institute, Baltimore, MD (2010).

[35] W.V. Dixon, Ed., *Cosmic Origins Spectrograph Instrument Handbook for Cycle 19*, Space Telescope Science Institute, Baltimore, MD (2010).

[36] E. P. Nelan, *Fine Guidance Sensor Instrument Handbook for Cycle 19*, Space Telescope Science Institute, Baltimore, MD (2010).

[37] L. Allen, R. Angel, J. D. Mangus, G. A. Rodney, R. R. Shannon, C. P. Spoelhof, *The Hubble Space Telescope Optical System Failure Report*. NASA Pub. TM-103443, NASA, Washington, DC (1990).

[38] *Ibid*. 14

[39] J. E. Krist, C. J. Burrows, "Phase Retrieval analysis of pre- and post-repair Hubble Space Telescope images," *Applied Optics* **34**(22), 4963 (1995).

[40] L. Allen, R. Angel, J. D. Mangus, G. A. Rodney, R. R. Shannon, C. P. Spoelhof, *The Hubble Space Telescope Optical System Failure Report*. NASA Pub. TM-103443, p.E-4, NASA, Washington, DC (1990).

[41] *Ibid*. 39

[42] *Ibid*.

[43] R. A. Brown & H.C. Ford, Eds., *A Strategy for Recovery, Report of the* HST *Strategy Panel*, Space Telescope Science Institute, Baltimore, MD (1991).

[44] P. Benvenuti, P. & E. Schreier, Eds., *Science with the Hubble Space Telescope, proceedings of ESO conference and workshop #44*, ESO, Chia Laguna, Sardinia, Italy (1992).

[45] R. L. White & R. J. Allen, Eds., *The Restoration of HST Images and Spectra; proceedings of a workshop held at the Space Telescope Science Institute, Baltimore, Maryland, 20-21 August 1990,* Space Telescope Science Institute, Baltimore, MD (1991).

[46] R. J. Hanisch & R. L. White, Eds., *The Restoration of HST Images and Spectra II,* Space Telescope Science Institute, Baltimore, MD (1994).

[47] http://hubble.nasa.gov/missions/intro.php

[48] http://www.spacetelescope.org/about/history/

[49] J. H. Crocker, "Engineering the COSTAR," *Optics & Photonics News* **4**(11), 22 (1993).

[50] J.E. Krist, "High Contrast Imaging with the Hubble Space Telescope: Performance and Lessons Learned," *Proc. SPIE* **5487**, 1284-1295 (2004).

[51] *OTA/FGS Thermal Control System Description and Operating Manual*, *TM-014140 / TN007-93S*, prepared for NASA/Goddard Space Flight Center by EER Systems Corp., Seabrook, MD (1993).

[52] M. Robberto, A. Sivaramakrishnan, J. J. Bacinski, D. Calzetti, J. E. Krist, J. W. MacKenty, J. Piquero, M. Stiavelli, "Performance of HST as an Infrared Telescope." *Proc. SPIE* **4013**, 386-393 (2000).

[53] M. Harwit, Ed., *Independent Science Review Team statement on the NICMOS Cryocooler*, http://www.stsci.edu/observing/nicmos_cryocooler_isr.html (1997).

[54] J. Grunsfeld, private communication (2011).

[55] http://hubble.nasa.gov/missions/sm4.php

[56] M. D. Lallo, R. B. Makidon, S. Casertano, J. E. Krist, "The Temporal Optical Behavior of HST: Focus, Coma, & Astigmatism History," *Proc. SPIE* **6270**, 62701N (2006).

[57] P.Y. Bély, H. Hasan, M. Miebach, "Orbital Variations in the Hubble Space Telescope," STScI Instrument Science Report SESD-93-16 (available from www.stsci.edu/hst/observatory/focus) (1993).

[58] D. DiNino, R. B. Makidon, M. Lallo, K. C. Sahu, M. Sirianni, S. Casertano, "HST Focus Variations with Temperature," STScI Instrument Science Report ACS-2008-03 (available from www.stsci.edu/hst/observatory/focus) (2008).

[59] C. Cox & S.-M. Niemi, "Evaluation of a temperature-based HST Focus Model," STScI Instrument Science Report TEL-2011-01 (available from www.stsci.edu/hst/observatory/focus) (2011).

[60] M. Lallo, R. van der Marel, C. Cox, G. Hartig, S.-M. Niemi, "HST Focus in SMOV4: Strategy for OTA adjustment & SI focus," STScI Instrument Science Report TEL-2010-02, Appendix A1 (available from www.stsci.edu/hst/observatory/focus) (2010).

[61] J. Anderson & I. R. King, "PSFs, Photometry, and Astrometry for the ACS/WFC," STScI Instrument Science Report, ACS-2006-01, (available from www.stsci.edu/hst/observatory/focus) (2006).

[62] J. D. Rhodes, R. J. Massey, J. Albert, N. Collins, R. S. Ellis, C. Heymans, J. P. Gardner, J.-P. Kneib, A. Koekemoer, A. Leauthaud, Y. Mellier, A. Refregier, J. E. Taylor, L. Van Waerbeke, "The Stability of the Point-Spread Function of the Advanced Camera for Surveys on the Hubble Space Telescope and Implications for Weak Gravitational Lensing," *Astrophys. J. Supp.* **172**(1) 203-218 (2007).

[63] R. B. Makidon, M. D. Lallo, S. Casertano, R. L. Gilliland, M. Sirianni, M., and J. E. Krist, "The Temporal Optical Behavior of the Hubble Space Telescope: The Impact on Science Observations," *Proc. SPIE* **6270**, 62701L (2006).

[64] J. L. Hershey, "Modelling HST Focal-Length Variations," STScI Instrument Science Report SESD-97-01 (available from www.stsci.edu/hst/observatory/focus) (1998).

[65] K. Kalinowski, private communication (2008).

[66] G. F. Hartig, L. Dressel, T. Delker, "WFC3 SMOV Programs 11424, 11434: UVIS Channel On-orbit Alignment," STScI Instrument Science Report WFC3-2009-45, p.6, (available from www.stsci.edu/hst/observatory/focus) (2009).

[67] G. F. Hartig, T. Delker, C. D. Keyes, "SMOV: COS NUV On-orbit Optical Alignment," STScI Instrument Science Report COS-2010-04, Sect. 3.2, (available from www.stsci.edu/hst/cos/documents/isrs) (2010).

[68] E. F. Arias, P. Charlot, M. Feissel, J.-F. Lestrade, "The extragalactic reference system of the International Earth Rotation Service, ICRS," *Astronomy & Astrophysics* **303** 604-608 (1995).

[69] B. M. Lasker, C. R. Sturch, B. J. McLean, J. L. Russell, H. Jenkner, M. M. Shara, "The Guide Star Catalog I - Astronomical Foundations and Image Processing." *Astronomical Journal* **99**(6), 2019-2078 (1990).

[70] L. Dressel, private communication (2005).

[71] B. M. Lasker, M. G. Lattanzi, B. J. McLean, B. Bucciarelli, R. Drimmel, J. Garcia, G. Greene, F. Guglielmetti, C. Hanley, G. Hawkins, V. G. Laidler, C. Loomis, M. Meakes, R. Mignani, R. Morbidelli, J. Morrison, R. Pannunzio, A. Rosenberg, M. Sarasso, R. L. Smart, A. Spagna, C. R. Sturch, A. Volpicelli, R. L. White, D. Wolfe, and A. Zacchei, "The Second-Generation Guide Star Catalog: Description And Properties," *Astronomical Journal* **136**, 735-766 (2008).

[72] M. Lallo, E. Nelan, E. Kimmer, C. Cox, S. Casertano, "Improving Hubble Space Telescope's Pointing & Absolute Astrometry," *Bulletin of the American Astronomical Society* **38**, 194 (2007).

[73] R. L. Gilliland, "Guiding Errors in 3-Gyro: Experience from WF/PC, WFPC2, STIS, NICMOS and ACS," STScI Instrument Science Report TEL-2005-02 (available from www.stsci.edu/hst/observatory/focus) (2005)

[74] *Ibid*.

[75] http://www.stsci.edu/hst/HST_overview/TwoGyroMode/documents/si_isrs.html

[76] J. D. Rhodes, R. Massey, J. Albert, J. E. Taylor, A. M. Koekemoer, A. Leauthaud, "Modeling and Correcting the Time-Dependent ACS PSF," in *2005 HST Calibration Workshop: Hubble after the transition to two-gyro mode, Proceedings of a workshop held at the Space Telescope Science Institute, Baltimore, Maryland, October 26-28, 2005* , A. M. Koekemoer, P. Goudfrooij, L. L. Dressel, Eds., p.21 (2006).

[77] J. Anderson & I. R. King, "PSFs, Photometry, and Astrometry for the ACS/WFC," STScI Instrument Science Report, ACS-2006-01, (available from www.stsci.edu/hst/observatory/focus) (2006).

[78] J. Biretta, M. McMaster, S. Baggett, S. Gonzaga, "Summary of WFPC2 SM97 Plans," STScI Instrument Science Report, WFPC2-97-03, (available from www.stsci.edu/instruments/wfpc2/Wfpc2_isr) (1997)

[79] R. A. Osten, D. Massa, A. Bostroem, A. Aloisi, C. Proffitt, "Updated Results from the COS Sensitivity Monitoring Program," STScI Instrument Science Report COS-2011-02, (available from www.stsci.edu/hst/cos/documents/isrs) (2011)

[80] M. Mutchler, Ed., *Workshop on Hubble Space Telescope CCD Detector CTE*, *Space Telescope Science Institute, Baltimore, Maryland, 31 Jan. – 1 Feb 2000*, (only available from www.stsci.edu/hst/acs/performance/cte_workgroup/cte_papers.html) (2000)

[81] A. Reiss & J. Mack, "Time Dependence of ACS WFC CTE Corrections for Photometry and Future Predictions", STScI Instrument Science Report ACS-2004-06 (available from www.stsci.edu/hst/acs/documents/isrs) (2004)

[82] J. Anderson & L. R. Bedin, "An Empirical Pixel-Based Correction for Imperfect CTE. I. HST's Advanced Camera for Surveys," *Publications of the Astronomical Society of the Pacific* **122**(895), 1035-1064 (2010)

[83] R. Massey, C. Stoughton, A. Leauthaud, J. Rhodes, A. Koekemoer, R. Ellis, E. Shaghoulian, "Pixel-based correction for Charge Transfer Inefficiency in the Hubble Space Telescope

Advanced Camera for Surveys," *Monthly Notices of the Royal Astronomical Society* **401**(1), 371-384 (2010)

[84] J. Townsend, "Hubble Space Telescope Multi-Layer Insulation Failure Review Board Results," Space Environment and Effects Program, Flight Experiments Workshop, June 23-25, 1998, Marshal Space Flight Center, Huntsville, Alabama,  (available at http://see.msfc.nasa.gov/fliexp_workshop/HST%20MLI_Jackie%20Townsend.pdf) (1998)

[85] A. S. Fruchter & R. N. Hook, "Drizzle: A Method for the Linear Reconstruction of Undersampled Images," *Publications of the Astronomical Society of the Pacific* **114**(792), 144-152 (2002)

[86] *Ibid*. 82

[87] *Ibid*. 83

[88] http://www.stsci.edu/hst/acs/software/CTE/

[89]  A. Fruchter, M. Sosey, W. Hack, L. Dressel, A. M. Koekemoer, J. Mack, M. Mutchler, N. Pirzkal, Eds., *The MultiDrizzle Handbook, version 3.0,* Space Telescope Science Institute, Baltimore, MD (2009)

[90] http://hla.stsci.edu/

[91] http://www.usvao.org/

[92] http://www.stsci.edu/institute/conference/calhst

[93] http://www.stsci.edu/institute/conference/cal97

[94] http://www.stsci.edu/institute/conference/cal02

[95] http://www.stsci.edu/institute/conference/cal05

[96] http://www.stsci.edu/institute/conference/cal10

[97] A. G. Riess, L. Macri, S. Casertano, H. Lampeitl, H. C. Ferguson, A. V. Filippenko, S. W. Jha, W. Li, R. Chornock, "A 3% Solution: Determination of the Hubble Constant with the Hubble Space Telescope and Wide Field Camera 3," *Astrophysical Journal* **730**(2), 119 (2011)

[98] "NASA's Hubble Finds Most Distant Galaxy Candidate Ever Seen in Universe," NASA Press Release, (www.nasa.gov/mission_pages/hubble/science/farthest-galaxy.html) (2011)

[99] *Ibid*. 3, p. 35

[100] D. J. Lindler, "Block Iterative Restoration of Astronomical Images from the Hubble Space Telescope," in *The Restoration of HST Images and Spectra; proceedings of a workshop held at the Space Telescope Science Institute, Baltimore, Maryland, 20-21 August 1990,* R. L. White & R. J. Allen, Eds., p.47, Space Telescope Science Institute, Baltimore, MD (1991).

AUTHOR BIOGRAPHY

Matthew Lallo is a Missions Systems Scientist at the Space Telescope Science Institute. Having a background in astronomy, he has worked on the Hubble project since prior to the telescope's deployment. He is currently involved with Hubble's successor, the James Webb Space Telescope, in the areas of commissioning, wavefront sensing and control, and overall observatory performance and characterization.

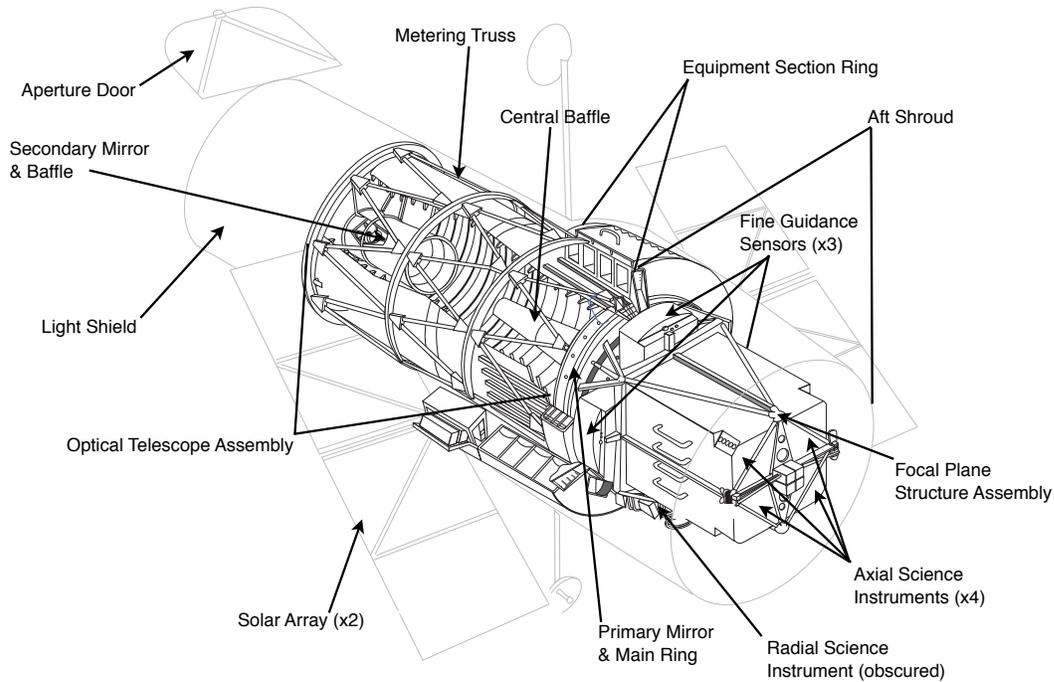

*Figure 1. HST Schematic showing items of interest within the telescope, spacecraft, and science instrument subsystems. Not labeled are the obscured 3 Rate Sensing Units, each containing two gyros, and 3 Fixed Head Star Trackers, which are all co-located on a single dimensionally stable tray lying aft of the Radial Science Instrument. The sunlit side is the top half of the cylindrical spacecraft (direction of the aperture door). In normal science operations sunlight does not encroach more than 20°-30° around onto the shadowed half, where the radial Science Instrument and the Star Trackers reside. Total length measured from aft end of the aft shroud to fore edge of the light shield is 13 m. (Figure adapted from STScI sources)*

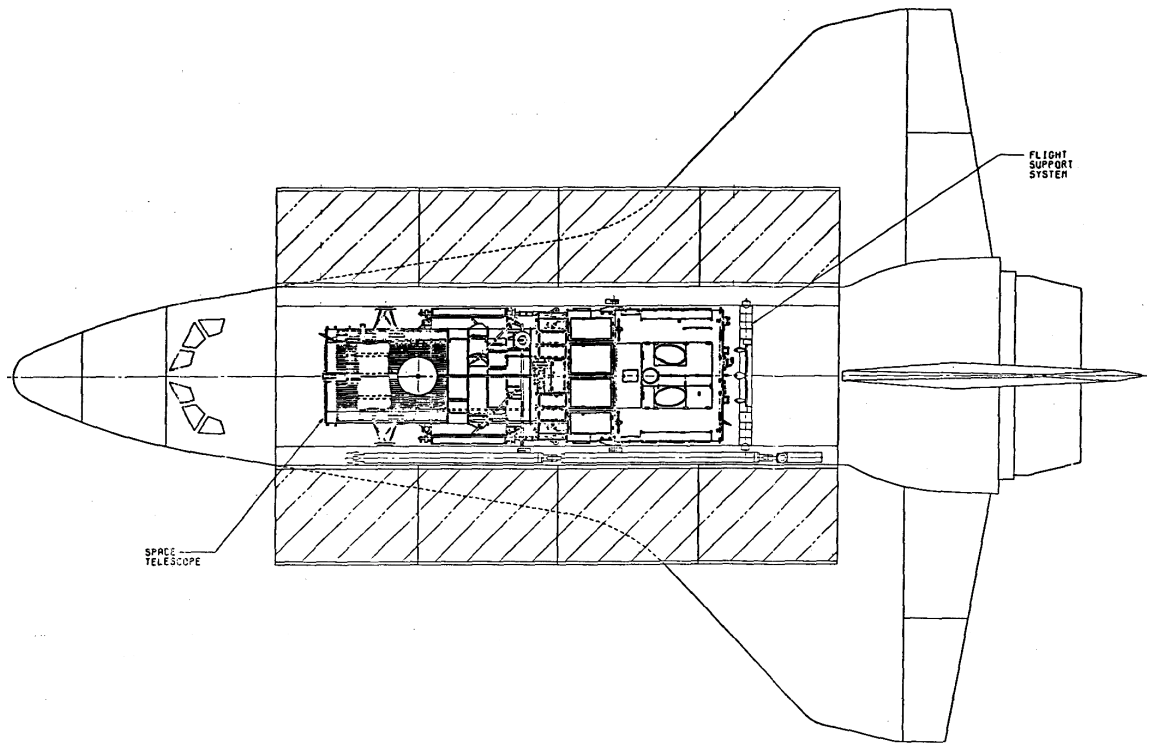

*Figure 2. HST & Shuttle Orbiter to scale, showing relative dimensions and approximate physical limitations on the diameter of the monolithic primary mirror. Folded deployable optics technologies, such as those used in the James Webb Space Telescope allow for large telescope diameters compared with launch vehicle payload capacities and dimensions. (Figure taken from [12])*

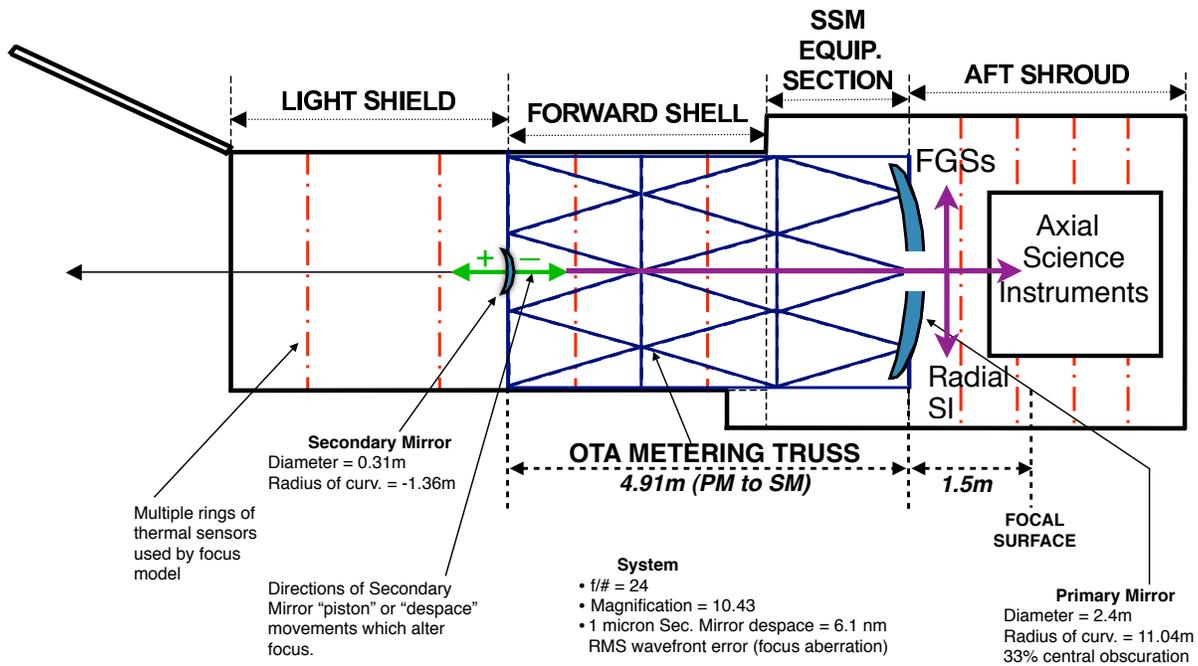

*Figure 3. Simplified HST schematic showing relevant optical quantities and overall physical layout. See also Table 1.*

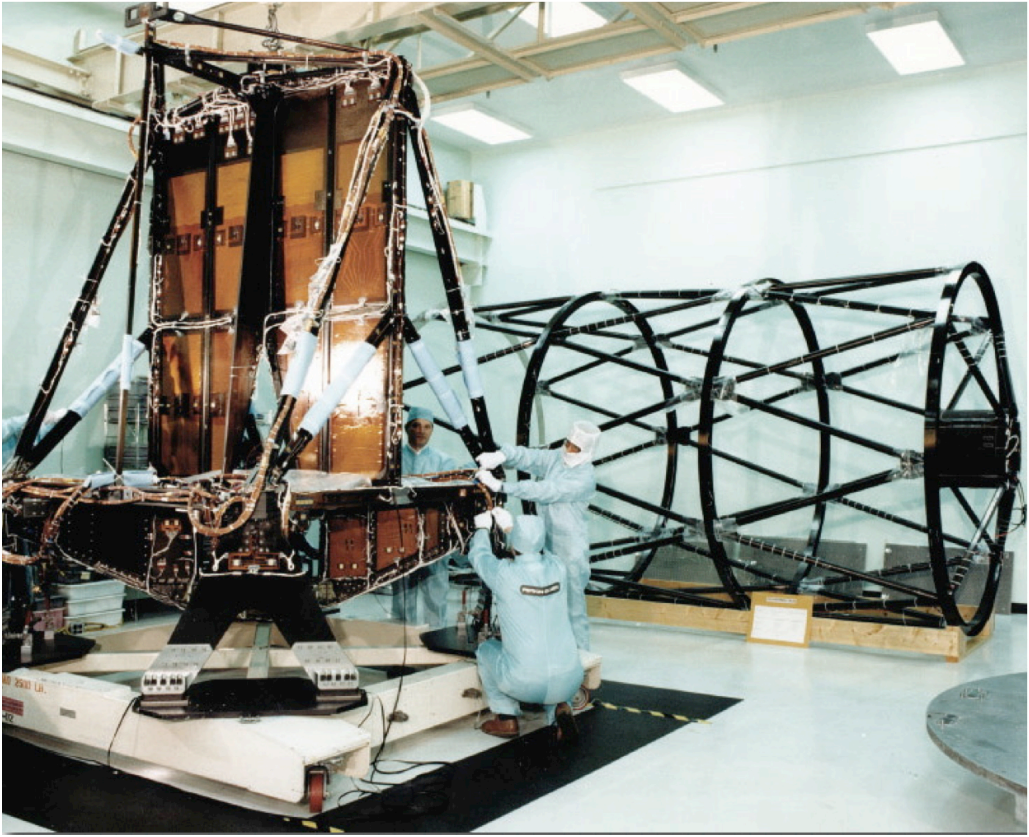

*Figure 4. The flight OTA Metering Truss and Focal Plane Structure Assembly (FPSA) during early stages of integration. The graphite epoxy truss retains the primary and secondary mirrors in very good alignment, resulting in no detectable non-axial motions. The truss has however steadily contracted longitudinally over the life of the mission necessitating periodic refocusing via secondary mirror actuation. In front of the truss, being tended to by technicians is the FPSA which houses the four "axial" Science Instruments, latched into place in each of the four rectangular quadrants, oriented vertically in this view. The FPSA has maintained the relative positions of the axial SIs tightly enough to be negligible compared with other source of dimensional changes such as moisture desorption within the instrument. [Image Perkin Elmer].*

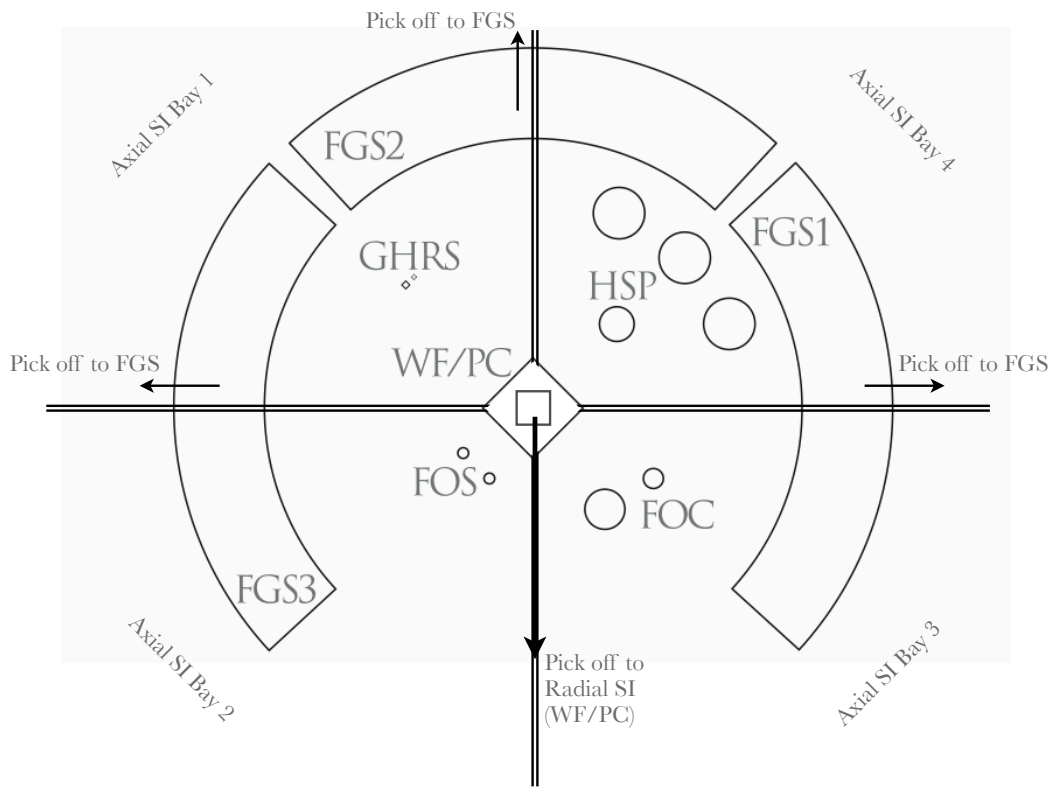

*Figure 5. The ~28 arcminute focal plane of HST as spatially shared by the 4 axial and 4 radial 1st generation instruments at the time of HST deployment. Each axial SI is a modular rectangular box roughly the dimensions of a phone booth which latches into the focal plane structure assembly in one of four bays. Entrance apertures very near the inner corners of the axial SI allow light from the OTA to pass into the instrument where it is then re-imaged or otherwise utilized. Sitting ahead of the fore end of the axial SIs are three arc-shaped 90° pick-off mirrors which feed the three Fine Guidance Sensors (radial instruments). These white-light shearing interferometers see the more astigmatic outer few arcminutes of the HST field of view. The fourth radial bay houses the radial SI which holds a pick-off mirror at the on-axis position at the center of the field of view. Figure 9 illustrates the replacement over the mission of instruments in each of the bays.*

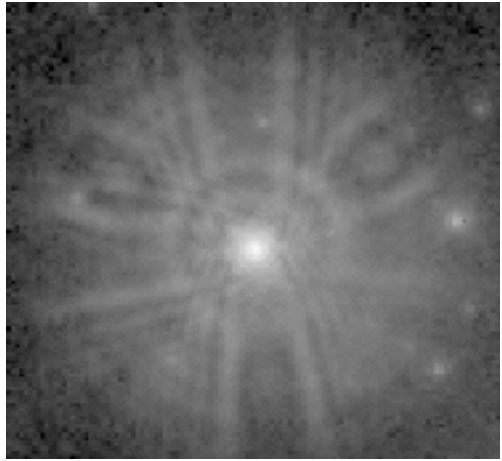

*Figure 6. Spherically aberrated WF/PC1 (PC) image. Only 15% of the energy was contained within the central 0.1". The remaining 85% was spread well into the PSF wings out to ~1.5", revealing structure attributable to the three primary mirror support pads, spaced 120° apart, and the secondary mirror support structures, spaced 90° apart and coincident with one of the pads. The tight ~0.1 arcsecond core was nonetheless unprecedented. (Image courtesy STScI)*

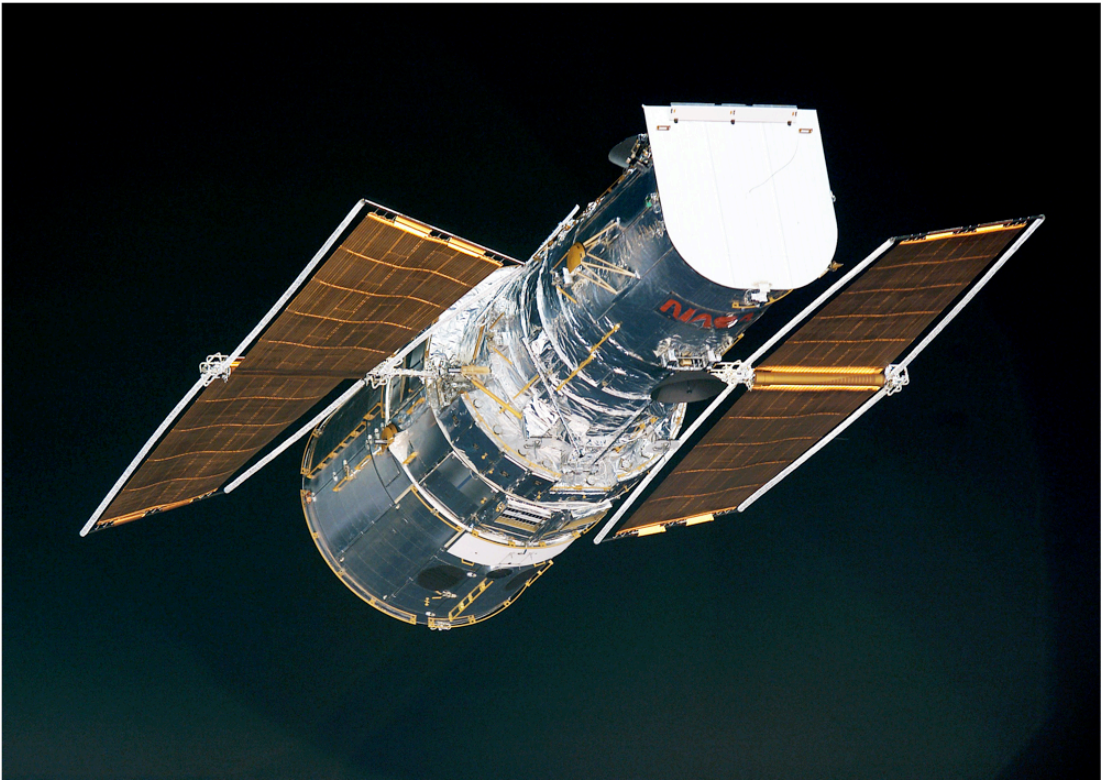

*Figure 7. Image of HST taken upon release from shuttle Endeavour at the completion of Servicing Mission 1, December 1993. New hardware visible are the 2nd generation solar arrays (with thermally insulating sleeve over bistems), and the white cylindrical-sectioned radiator of the 2nd generation radial camera, WFPC2. (Image courtesy NASA).*

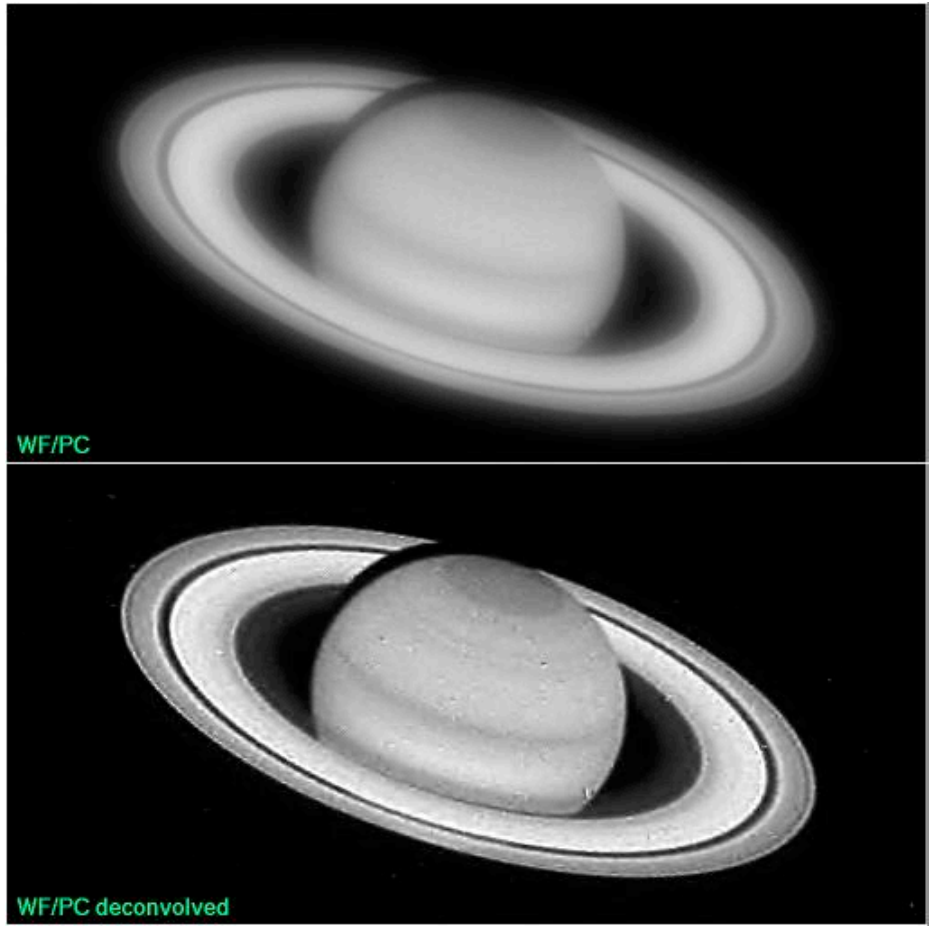

*Figure 8. Deconvolution example of early science. Saturn as imaged with WFPC1 F718 filter, original and reconstructed using a block iterative restoration algorithm described in [100]. (Image courtesy Don Lindler[100]).*

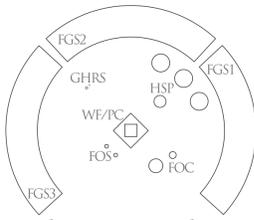
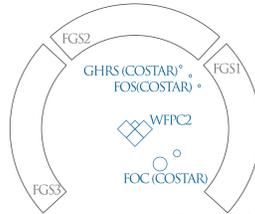
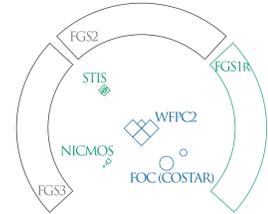
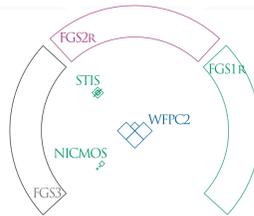
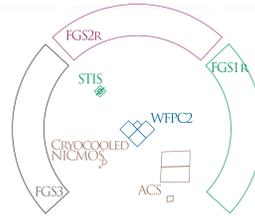
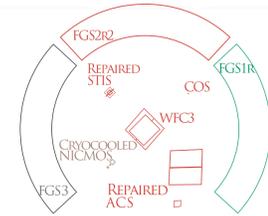

Figure 9. Evolution of HST via human servicing missions. Changes to the telescope's science instrument complement is highlighted by color-coding. The current incarnation of HST can be seen to contain science equipment from SM4, SM3B, SM2, and (in the case of FGS3) the original deployment.

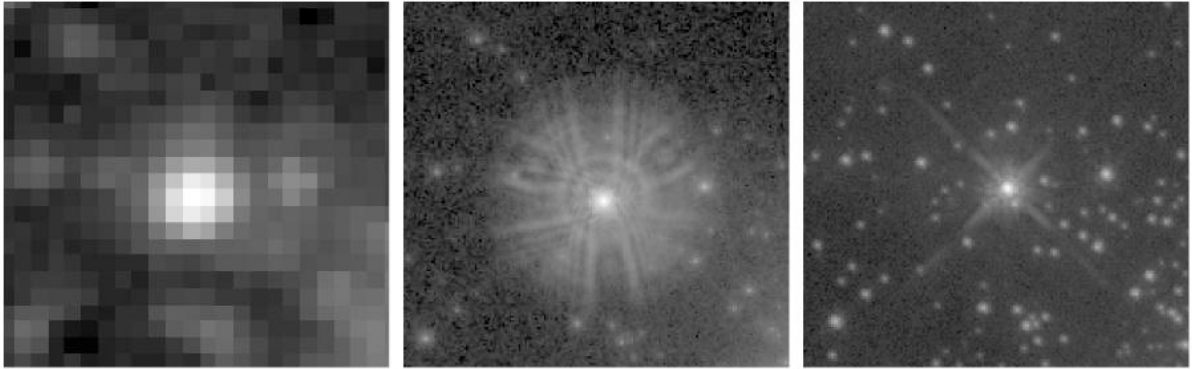

*Figure 10. A comparison of PSFs from left to right: ground image at 0.6" resolution, spherically aberrated WF/PC1 image (described in Fig. 6), and WFPC2 image after 1$^{st}$ servicing mission, with PSF characteristics meeting and exceeding requirements. [Image courtesy STScI].*

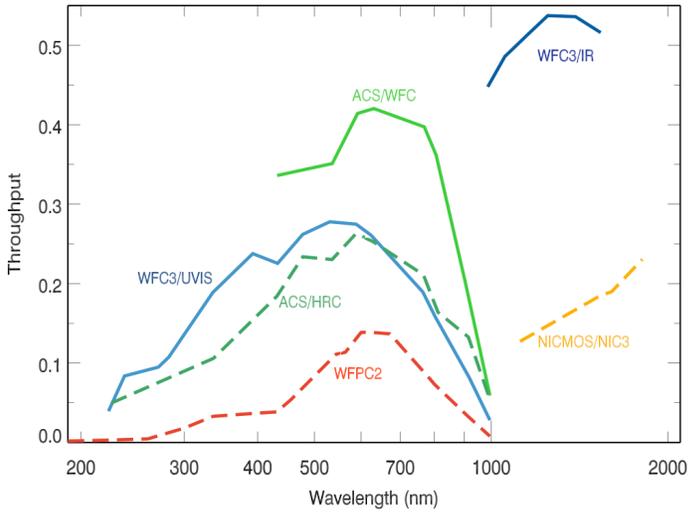 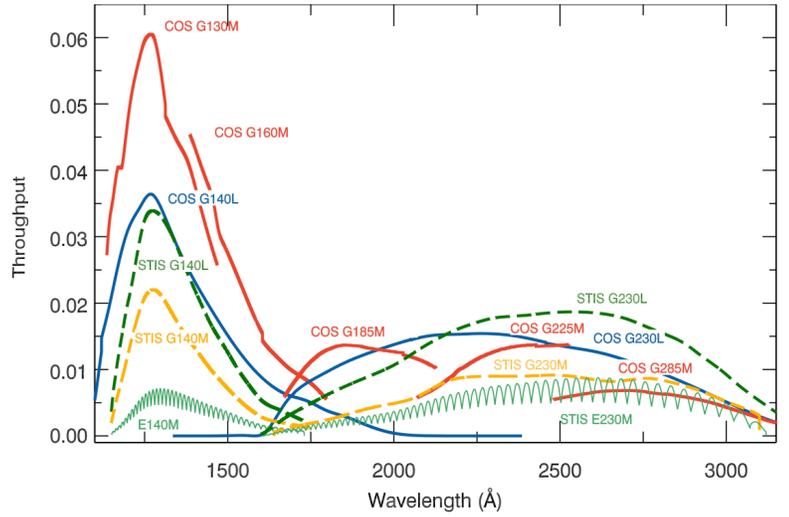

*Figure 11: Left pane: Total fractional system throughput (OTA included) for the present HST cameras' (left) and spectrographs (right). The post-SM4 instruments, WFC3 and COS, provide significant increases in IR imaging sensitivity and far-UV spectral sensitivity, respectively. (Figure adapted from STScI sources).*

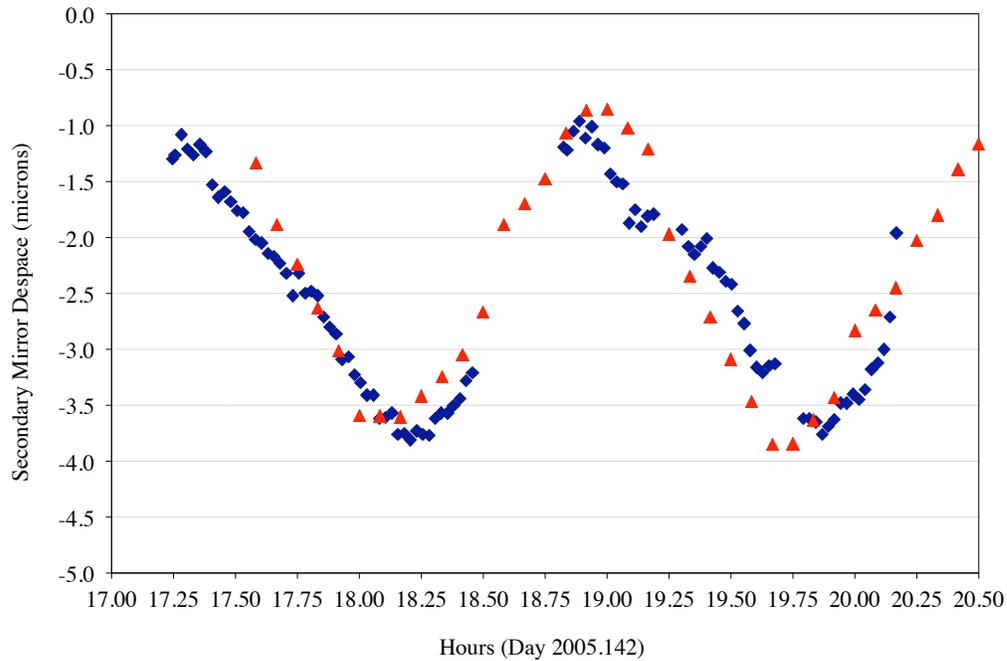

*Figure 12: A typical change in focus over an HST orbit is seen here to be ~3 microns of secondary mirror despace, or ~18 nanometers rms wavefront error (~$\lambda/28$ in V-band). The diamond-shaped points are measurements of a single PSF in individual exposures taken with the Advanced Camera for Surveys' High Resolution Channel. The stellar image is relatively well sampled with pixels spanning 25 milliarcseconds. Parametric phase retrieval was performed on the PSF, and the best fit focus (Zernike term $Z_4$) was expressed in terms of microns of secondary mirror despace. The triangular points are model values. This empirically derived model expresses focus changes as a function of temperatures as telemetered from a number of stations throughout the observatory. The temperatures which correlate closest with orbital focus swings are taken from points around the telescope's aft light shield area which is just fore of the secondary mirror support structure. Of note from this plot is that 1.) the phase retrieval measurement error is quite good, with the noise in the $Z_4$ determinations being at the nanometer level (1 micron at the secondary gives ~6 nanometers of rms wavefront error) and 2.) the clearly measurable real changes at this level illustrate the image quality's sensitivity to changes well under the "diffraction limit", i.e. focus wavefront errors of ~ $\lambda/500$ are detectable.*

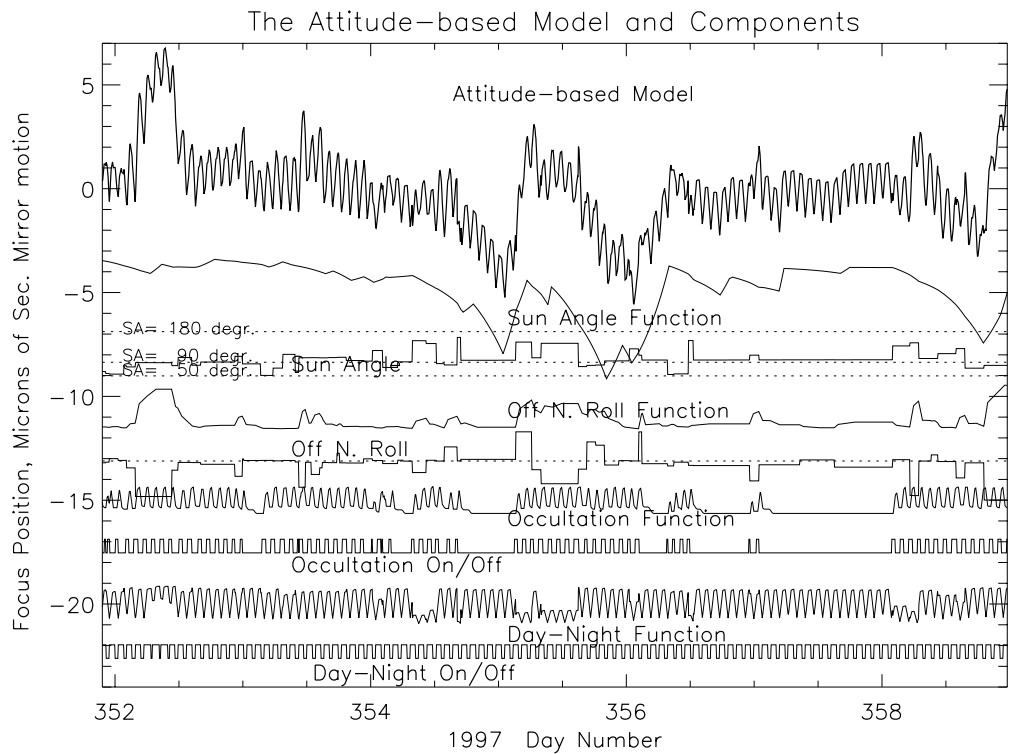

*Figure 13: One of the focus models (top curve) plotted over a 1 week period along with the HST attitude-related parameters found to have the highest correlation with focus changes. Sun angle is the angle of the sun from the HST line of sight (180° is an antisun pointing). This is known to strongly affect temperatures that determine secondary mirror despace. "Off-N. Roll" is the roll angle with respect to the sun (0 when sun is in the direction of aperture door hinge) Earth occultation affects temperatures due to IR heating from the earth, as does obviously sunlight during day/night changes. (Figure taken from [64]).*

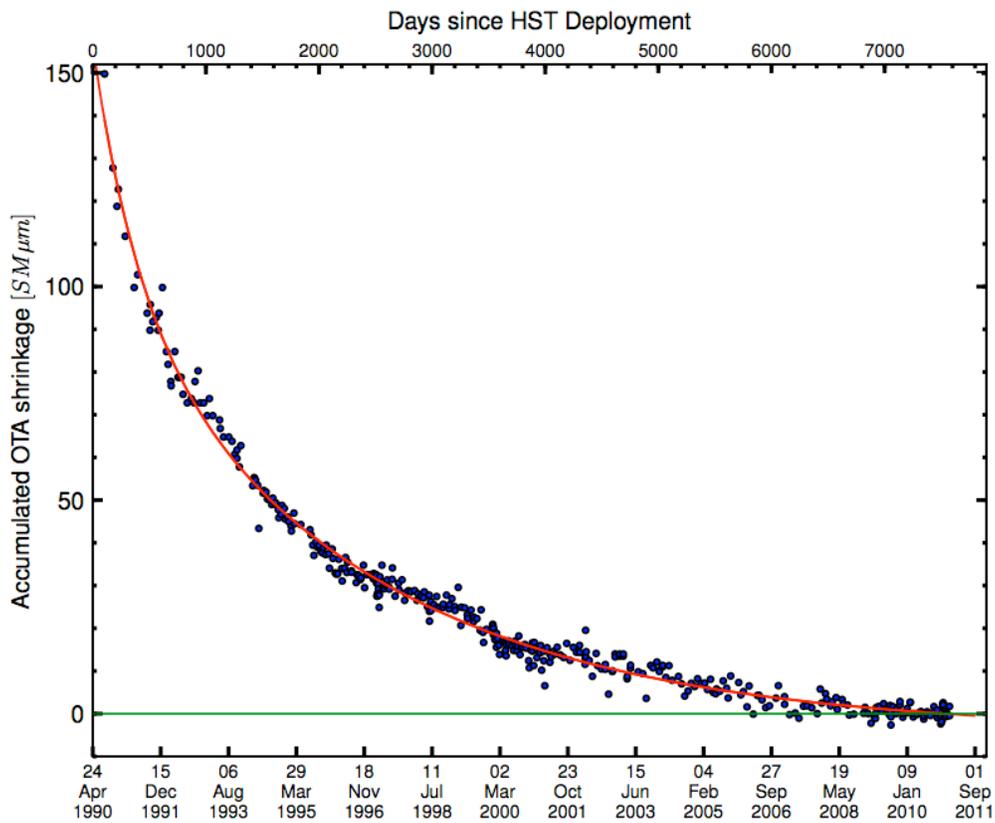

*Figure 14: Total shrinkage of the OTA Metering Truss as inferred from phase retrieval measurements of the focus aberration over the mission life. Total shrinkage is $3 \times 10^{-5}$ the truss length or 0.003% over 20 years.*

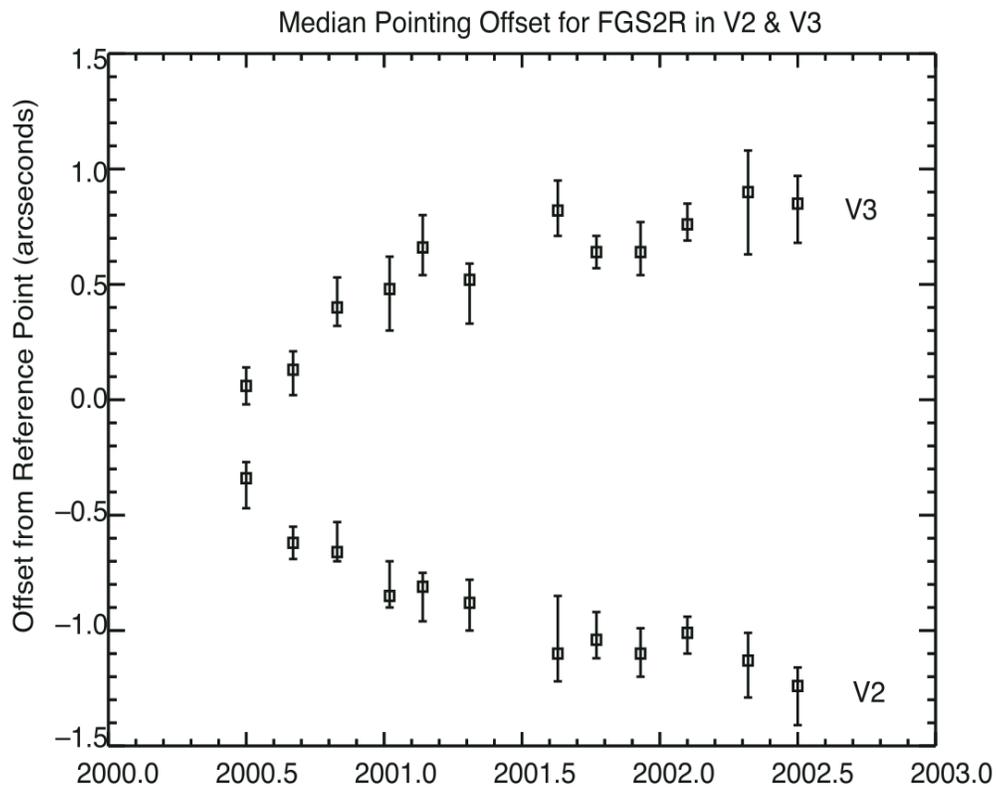

*Figure 15:Measured trend in FGS2R's location in the focal plane tangent frame. Outgassing in the years following its installation in SM3A is the suspected cause. All SIs and FGSs have exhibited this type of behavior, calling for more frequent calibration during the time to ensure pointing accuracy. (Figure adapted from [70]).*

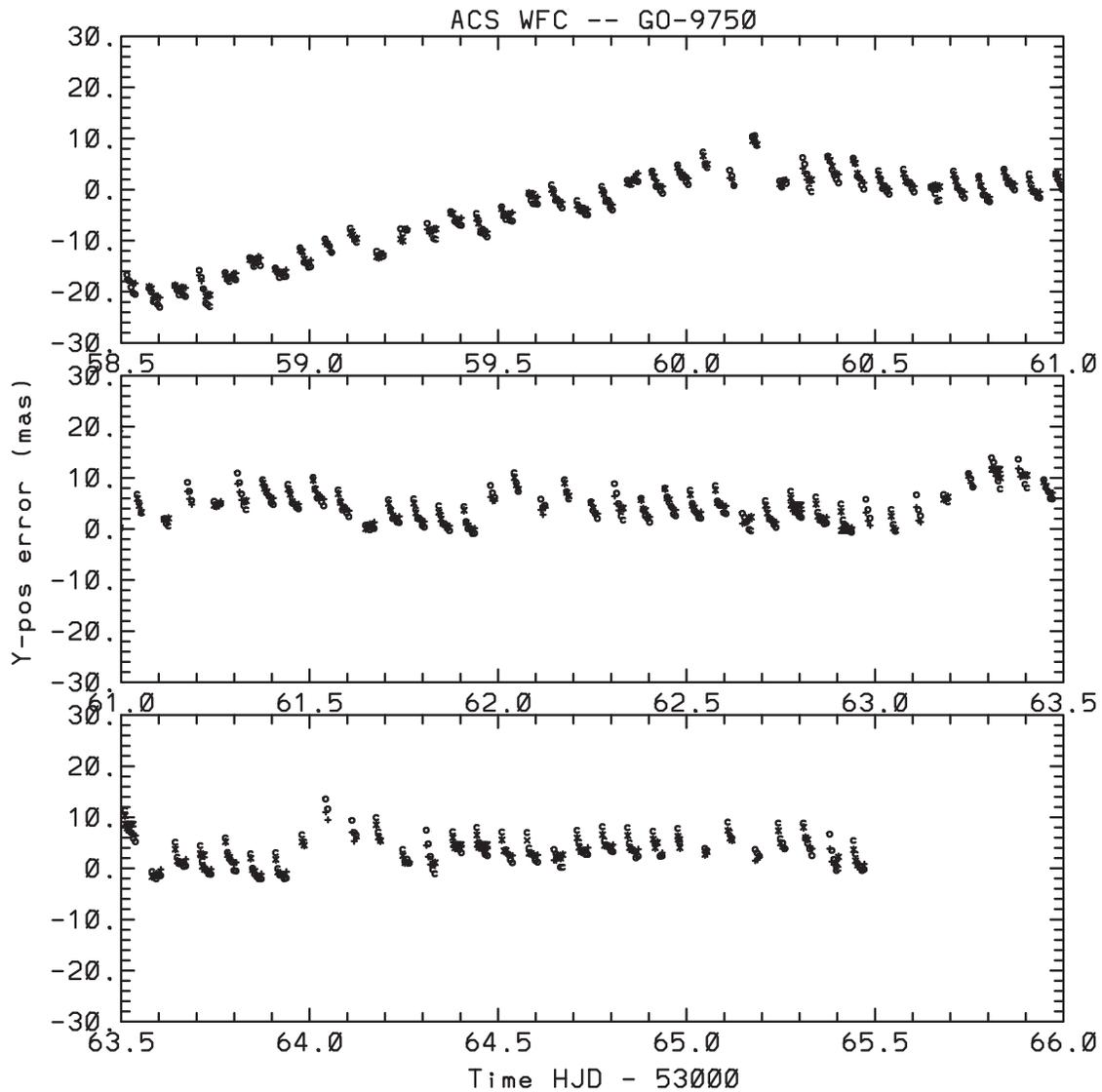

*Figure 16: Pointing drift as measured in ACS WFC over 7 days. Time axis is wrapped from one pane to the next. Relative pointing was determined for each exposure in a long science program. Precise relative pointing determinations of ~1* mas *were enabled by measuring pixel positions of a large number of stars. The smallest scale structure is the drift over an orbit. Pointing shifts at guidestar re-acquisitions are obvious as the discontinuities. The long-term trend that stabilizes as temperatures equilibrate at a given attitude can be seen. (Figure adapted from [73]).*